\documentclass[showpacs,aps,prc,nofootinbib,showkeys,twocolumn]{revtex4-1}

\usepackage{graphicx}
\usepackage{bm}
\usepackage{amssymb}
\usepackage{amsmath,latexsym}
\usepackage[usenames]{color}
\usepackage[dvipsnames]{xcolor}
\usepackage{subfigure}
\usepackage{slashed}
\usepackage{multirow,array}
\usepackage{mathtools}
\usepackage{mathrsfs}
\usepackage[colorlinks=true,linktocpage=true]{hyperref}
\usepackage[utf8]{inputenc}
\usepackage{hyperref}
\hypersetup{colorlinks,linkcolor={blue},citecolor={blue},urlcolor={magenta}}
\usepackage{lipsum}
\usepackage{dsfont}
\usepackage{soul}
\newcommand{\be}{\begin{equation}}
\newcommand{\ee}{\end{equation}}
\newcommand{\bea}{\begin{eqnarray}}
\newcommand{\eea}{\end{eqnarray}}

\newcommand{\diag}{\mathop{\mathrm{diag}}}

\newcommand{\eq}{\text{eq}}

\newcommand{\munu}{{\mu\nu}}
\newcommand{\trento}{{\sc T}$_\text{\sc R}${\sc{ENTo}}}
\newcommand{\ds}{s}

\newcommand{\del}{\partial}

\newcommand{\px}{p^{\langle\mu\rangle}}

\newcommand{\up}{u {\,\cdot\,} p}

\newcommand{\pOp}{(p \cdot \Omega \cdot p)}

\newcommand{\ene}{\mathcal{E}}

\newcommand{\PL}{\mathcal{P}_L}
\newcommand{\Pperp}{\mathcal{P}_\perp}
\newcommand{\Peq}{\mathcal{P}_\text{eq}}
\newcommand{\bs}{\begin{subequations}}
\newcommand{\es}{\end{subequations}}
\newcommand{\beal}{\begin{align}}
\newcommand{\enal}{\end{align}}

\newcommand{\M}{\mathcal{M}}

\newcommand{\J}{\mathcal{J}}
\newcommand{\N}{\mathcal{N}}
\newcommand{\A}{\mathcal{A}}
\newcommand{\B}{\mathcal{B}}

\newcommand{\piperp}{\pi_\perp}
\newcommand{\Wperp}{W_{\perp z}}

\begin{document}

\title{Modified equilibrium distributions for Cooper--Frye particlization}
\date{\today}
\author{M.~McNelis}
\affiliation{Department of Physics, The Ohio State University, Columbus, OH 43210, USA}
\author{U.~Heinz}
\affiliation{Department of Physics, The Ohio State University, Columbus, OH 43210, USA}

\begin{abstract}
We introduce a positive definite single-particle distribution that is suitable for describing the transition from a macroscopic hydrodynamic to a microscopic kinetic description during the late stages of heavy-ion collisions in the presence of moderately large viscous corrections. The modified equilibrium distribution function can be constructed with hydrodynamic input from either relativistic viscous fluid dynamics or anisotropic fluid dynamics. We test the modified equilibrium distribution's hydrodynamic output for a stationary hadron resonance gas subject to either shear stress, bulk pressure, or baryon diffusion current at a given freeze-out temperature and baryon chemical potential. While it does not reproduce all components of the net baryon current and energy-momentum tensor exactly, it significantly improves upon the customary linearized approximations for the non-equilibrium correction $\delta f_n$ which typically lead to unphysical negative distribution functions at large particle momenta. A comparison of particle spectra and $p_T$--differential elliptic flow coefficients from the Cooper--Frye formula computed with the modified equilibrium distribution and with linearized $\delta f_n$ corrections is presented, for two different (2+1)--dimensional hypersurfaces corresponding to central and non-central Pb+Pb collisions at the Large Hadron Collider (LHC).
\end{abstract}

\pacs{12.38.Mh, 25.75.-q, 24.10.Nz, 52.27.Ny, 51.10.+y}
\keywords{Cooper--Frye formula, ultrarelativistic heavy-ion collisions, relativistic kinetic theory, 14--moment approximation, Chapman--Enskog expansion, modified equilibrium distribution}

\maketitle

\section{Introduction}
\label{sec1}

Among a large variety of computational approaches describing the evolution of ultra-relativistic heavy-ion collisions, hybrid models have had the greatest success in describing a wide variety of hadronic observables simultaneously~\cite{Song:2010mg, Schenke:2011bn, Gale:2012rq, Gale:2013da, Shen:2014lye, Ryu:2015vwa, Bernhard:2016tnd, Bernhard:2018hnz}. Hybrid models include (at least) a relativistic viscous hydrodynamics module describing the spacetime evolution of the early hot and dense quark-gluon plasma (QGP) stage of the fireball created in the collision, followed by a hadron cascade ``afterburner'' that follows the many-particle system of hadronic resonances created from the QGP during the process of ``hadronization'' microscopically to its final ``kinetic freeze-out''. While experimental data strongly suggest that the QGP stage evolves as a near-perfect liquid~\cite{Heinz:2001xi, Romatschke:2007mq, Song:2010mg, Schenke:2010rr, Schenke:2010nt, Denicol:2012cn, Heinz:2013th}, its transport properties cannot be measured directly, nor can they (yet) be reliably computed from first principles. Rather, they must be inferred using various experimental probes, the most abundant of which (and therefore most precisely measured) are the soft-momentum hadrons produced after the quark-gluon plasma has cooled down below its pseudo-critical temperature. 

Modeling the emission of these hadrons is therefore a key component in any hybrid model~\cite{Shen:2014vra}. However, the exact spatial and momentum configurations of these emitted hadrons are not well understood since their distribution function is governed by microscopic kinetic theory while the preceding quark-gluon plasma phase is treated as a strongly-coupled fluid. The dynamical process in which the quark-gluon plasma converts to hadrons, known as hadronization, is a highly complex, unresolved problem. Instead, hybrid models bridge the transition between these two different phases of QCD matter with a particlization model where the strongly-coupled quark-gluon plasma is evolved hydrodynamically (i.e. macroscopically, with minimal microscopic input such as the equation of state (EoS) which can be obtained from lattice QCD~\cite{Borsanyi:2010bp, Borsanyi:2010cj, Bazavov:2011nk, Borsanyi:2013bia, Bazavov:2014pvz} and parametrized transport coefficients) until hadronization is complete, followed by an instant conversion to weakly-interacting hadrons on a hypersurface $\Sigma(x)$~\cite{Huovinen:2012is}. The conversion between different sets of microscopic degrees of freedom relies on the assumption that both viscous hydrodynamics and kinetic theory are simultaneously valid on this surface $\Sigma(x)$.

Requiring that the energy, momentum and charges of the system are conserved during this ``particlization process'', the Cooper--Frye formula gives the particle spectrum for hadron species $n$ as~\cite{PhysRevD.10.186}
\be
\label{eq:CooperFrye}
  E_n \frac{dN_n}{d^3p} = \frac{1}{(2\pi \hbar)^3}\int_\Sigma p \cdot d^3\sigma(x) \, f_n(x,p)\,,
\ee
where $f_n$ is its distribution function in phase-space~\cite{PhysRevD.10.186}. For a locally equilibrated hadron resonance gas this distribution function takes the form 
\be
\label{eq:feq}
 f_{\mathrm{eq},n}(x,p) = \frac{g_n}{\exp\Bigl[\dfrac{p \cdot u(x)}{T(x)} - \alpha_n(x)\Bigr] + \Theta_n} \,,
\ee
where $g_n$ is the spin degeneracy factor, $u^\mu(x)$ is the fluid velocity, $T(x)$ is the temperature, $\alpha_n(x) = \mu_n(x)/T(x)$ is the chemical potential-to-temperature ratio of species $n$, and $\Theta_n \in [1,-1]$ accounts for the quantum statistics of fermions and bosons, respectively. In this work, we only consider the baryon chemical potential $\mu_B$ and write $\alpha_n = b_n \alpha_B$ where $b_n$ is the baryon number of species $n$. 

Due to dissipative effects in the preceding hydrodynamic evolution of the fluid, the distribution function is generally out of local-equilibrium on the conversion surface. One writes $f_n(x,p)=f_{\mathrm{eq},n}(x,p)+\delta f_n(x,p)$ where $\delta f_n(x,p)$ encodes the deviation from local equilibrium. The temperature and chemical potential for the equilibrium part are obtained from the energy and net-baryon densities by Landau matching. Deviations from local equilibrium are described in the hydrodynamic stage by dissipative corrections to the energy-momentum tensor $T^{\mu\nu}(x)$ and the net baryon current $J_B^\mu(x)$. These so-called ``dissipative flows'' are the baryon diffusion current $V_B^\mu(x)$ (a spatial vector in the local fluid rest frame (LRF), characterized by three degrees of freedom), the shear stress tensor $\pi^{\mu\nu}(x)$ (a symmetric traceless rank-2 tensor with only spatial components in the LRF, characterized by five degrees of freedom) and the bulk viscous pressure $\Pi(x)$ (adding a single scalar degree of freedom). In kinetic theory these dissipative flows are defined as momentum moments of the off-equilibrium part $\delta f_n$ of the distribution functions, summed over all hadron species $n$. At each spacetime point $x$ their values provide altogether nine constraints on the momentum dependences of {\it all of the off-equilibrium corrections} $\delta f_n(x,p)$, $n=1,\dots,N_R$, but obviously these are not sufficient to fully determine them.    

In this sense, the particlization problem is ill-posed. In contrast to a full solution of the underlying kinetic equations, the hydrodynamic output provides only limited information about $\delta f_n(x,p)$. To fully specify the form of $f_n(x,p)$ in the Cooper--Frye formula (\ref{eq:CooperFrye}), additional guiding principles are needed to make optimal use of the hydrodynamic constraints. One well-known scheme is Grad's 14--moment approximation~\cite{CPA:CPA3160020403} in which the momentum dependence of $\delta f_n$ is expanded up to second order in the momenta, with coefficients matched to the 14 hydrodynamic moments of $\{f_n\}$, $T^{\mu\nu}$ and $J^\mu_B$, and assuming that all particle species $n$ have the same expansion coefficients. Another is the first-order Chapman--Enskog expansion of the Boltzmann equation in the relaxation time approximation (RTA), with a common relaxation time for all hadron species~\cite{chapman1990mathematical, Anderson_Witting_1974}. Both schemes treat $\delta f_n$ as a linear perturbation of $f_n(x,p)$, ignoring higher-order terms when matching to the hydrodynamic constraints. Unfortunately, these linear corrections can turn the total distribution functions negative at high momenta, especially for large dissipative flows. In particular, when particlization is performed in close proximity to the quark-hadron phase transition, dissipative corrections related to bulk viscosity (which peaks near the pseudocritical temperature) tend to be large, and a linearized approach can no longer be trusted~\cite{Denicol:2009am}.

In the past decade, work has been done on constructing a non-equilibrium distribution function suitable for Cooper--Frye particlization that does not rely on a linearized expansion scheme. In the original work by Pratt and Torrieri \cite{Pratt:2010jt}, dissipative perturbations are added to the ``auxiliary fields'' $T(x)$, $u^\mu(x)$ and $\alpha_B(x)$ in the Boltzmann factor, effectively transforming the local-equilibrium distribution \eqref{eq:feq} into a quasi-equilibrium distribution. The resulting modified equilibrium distribution is given by the formula
\be
\label{eq:feqmod_original}
  f_{\eq,n}^\text{PT} = 
  \frac{\mathcal{Z}_n g_n}
       {\exp\biggl[\dfrac{\sqrt{{\bm{p}'^2{+}m_n^2}}}{T {+} \delta T} - b_n(\alpha_B {+} \delta \alpha_B)\biggr] + \Theta_n}\,,
\ee
where $\delta T = \lambda_T \Pi$ and $\delta\alpha_B = \lambda_{\alpha_B} \Pi$ are bulk viscous corrections to the effective temperature and chemical potential, while $p^{\,\prime}_i = - X_i \cdot p^{\,\prime}$ are modified local-rest-frame (LRF) momentum components, with $X^\mu_i = (X^\mu, Y^\mu, Z^\mu)$ being the spatial basis vectors (i.e. the four-vectors that reduce in the LRF to the directional unit vectors along the $x$, $y$, $z$ axes). The fluid velocity perturbation $\delta u^\mu$ is encoded in the momentum space transformation
\be
\label{eq:rescale}
  p_i = A_{ij} p^{\,\prime}_j \,,
\ee
where $p_i = - X_i \cdot p$ are the usual LRF momentum components and
\be
\label{eq:Aij}
  A_{ij} = (1 {+} \lambda_\Pi \Pi)\delta_{ij} + \lambda_\pi \pi_{ij}
\ee
is a symmetric matrix that deforms the momentum space linearly with the bulk viscous pressure $\Pi(x)$ and LRF shear stress tensor $\pi_{ij}(x) = X^\mu_i X^\nu_j \pi_\munu(x)$.\footnote{%
    The scalar coefficients $\lambda_T$, $\lambda_{\alpha_B}$, $\lambda_\Pi$ and $\lambda_\pi$ are functions of the temperature and chemical potential ($T$, $\alpha_B$).}$^,$\footnote{%
    Ref.~\cite{Pratt:2010jt} did not consider a nonzero baryon diffusion current $V_B^\mu(x)$.} 
The normalization factor 
\be
\label{eq:Z_PT}
 \mathcal{Z}_n = \frac{1}{{\det}A}
\ee
rescales the distribution function such that its particle density agrees with that of a local-equilibrium distribution with temperature $T{+}\delta T$ and chemical potential $\alpha_B{+}\delta\alpha_B$.

The modified equilibrium distribution  (\ref{eq:feqmod_original}) is constructed such that its momentum moments reproduce all components of $J_B^\mu$ and $T^\munu$ at first order in the (small) dissipative corrections. As the dissipative flows $\Pi$ and $\pi_{ij}$ become larger, however, the mismatch between the hydrodynamic input and the output when re-computing them as moments of the modified distribution function (\ref{eq:feqmod_original}) increases. This raises the question whether the modifications to the local-equilibrium distribution can minimize this mismatch even for moderately large viscous corrections. Tests conducted in Ref.~\cite{Pratt:2010jt} for the case of vanishing bulk viscous pressure ($\Pi=0$) showed that Eq.~\eqref{eq:feqmod_original} accurately reproduces the target shear stress as well as the energy and charge densities even for moderately large shear stress modifications. However, when repeated for moderately large bulk viscous pressures, this time setting the shear stress to $\pi_{ij} = 0$, the modified equilibrium distribution's hydrodynamic output strongly deviated from the input values. This suggests that the formula \eqref{eq:feqmod_original} can be further improved.

To reduce the errors associated with bulk viscous pressure, a variant of the Pratt--Torrieri modified equilibrium distribution has been developed and recently implemented in a Bayesian model parameter extraction of heavy-ion collision simulations of Pb+Pb collisions at LHC energies ($\sqrt{s_\text{NN}} = 2.76$ and 5.02 TeV)~\cite{Bernhard:2018hnz}. A different type of quasi-equilibrium distribution, developed much earlier but with somewhat similar goals in mind, is the anisotropic distribution~\cite{Romatschke:2003ms} on which anisotropic fluid dynamics is based \cite{Martinez:2010sc}. Although its analytic form has similarities with Eq.~\eqref{eq:feqmod_original}, the relation between the modification parameters and the dissipative flows is made nonlinear in order to capture the large pressure anisotropies present in rapidly longitudinally expanding systems \cite{Heinz:2015gka}. Anisotropic fluid dynamical simulations have been interfaced with particlization models based on the Romatschke--Strickland anisotropic distribution~\cite{Romatschke:2003ms} to describe, with considerable success, the particle spectra of Pb+Pb collisions at LHC energies ($\sqrt{s_\text{NN}}=2.76$ TeV) and Au+Au collisions at RHIC energies ($\sqrt{s_\text{NN}} = 200$ GeV)~\cite{Alqahtani:2017tnq, Almaalol:2018gjh}. A further generalization of this approach has been proposed in Ref.~\cite{Nopoush:2019vqc}. 

In this paper we present an improved formulation of the modified equilibrium distribution \eqref{eq:feqmod_original}. The perturbations made to the exponential argument of the local-equilibrium distribution \eqref{eq:feq} are derived systematically from the gradient expansion of the RTA Boltzmann equation. These modifications, which are proportional to the hydrodynamic gradients, are constructed in such a way that the modified equilibrium distribution reduces to the first-order RTA Chapman--Enskog expansion in the limit of small viscous corrections. We test its ability to reproduce the net baryon current and energy-momentum tensor of a stationary hadron resonance gas subject to a variety of dissipative flows at chemical freeze-out, including moderately large bulk viscous pressures and baryon diffusion currents. We further apply the same method to also modify a Romatschke--Strickland anisotropic distribution \cite{Romatschke:2003ms} at leading order (i.e. with only nonzero diagonal elements in the momentum deformation matrix $A_{ij}$) in such a way that it can approximately capture the smaller off-diagonal components of the energy-momentum tensor. We demonstrate the potential use of our modified equilibrium (anisotropic) distribution by computing the particle spectra from a longitudinally boost-invariant Pb+Pb collision and comparing them to those computed with the linearized $\delta f_n$ corrections as well as the modified equilibrium distribution derived in Ref.~\cite{Bernhard:2018hnz}.

The paper is structured as follows: in Sec.~\ref{sec2} we review the linearized viscous corrections to the hadronic distributions given by the 14--moment approximation and RTA Chapman--Enskog expansion. In Sec.~\ref{sec3} we derive the modified equilibrium distribution from the relaxation time approximation and test the reproduction of $J_B^\mu$ and $T^\munu$ for a stationary hadron resonance gas. In Sec.~\ref{sec4}, we repeat the procedure outlined in Sec.~\ref{sec3} but starting with a leading order anisotropic distribution. In Sec.~\ref{sec5} we compute the particle spectra from a (2+1)--dimensional viscous hydrodynamic simulation of Pb+Pb collisions at LHC energies ($\sqrt{s_\text{NN}} =2.76$ TeV) using the Cooper--Frye formula and compare the differences between the modified distributions and linearized $\delta f_n$ corrections. 

\section{Linearized viscous corrections} 
\label{sec2}
\subsection{14--moment approximation}
\label{sec2a}

The 14--moment approximation is a moments expansion of the distribution function around $f_{\eq,n}$ that is truncated to the 14 lowest momentum moments ($p^\mu$, $p^\mu p^\nu$)~\cite{CPA:CPA3160020403}. This approximation assumes that the distribution function can be adequately characterized by just its hydrodynamic moments (i.e. $\int_p p^\mu f_n$, $\int_p p^\mu p^\nu f_n$) ~\cite{Denicol:2012cn}. For a multi-component gas with nonzero baryon chemical potential, the 14--moment approximation reads~\cite{Monnai:2009ad} 
\be
\label{eq:14_moment}
  \delta f^{14}_n = f_{\eq,n} \bar{f}_{\eq,n} \left( b_n c_\mu p^\mu + c_\munu p^\mu p^\nu \right)\,,
\ee
where $\bar{f}_{\eq,n} = 1 - g_n^{-1} \Theta_n f_{\eq,n}$. For simplicity, we follow common practice and take the expansion coefficients $c_\mu$ and $c_\munu$ to be species-independent. To solve for the coefficients, one rewrites Eq.~\eqref{eq:14_moment} in irreducible form, 
\be
\label{eq:14_moment_irreducible}
\begin{split}
  \delta f^{14}_n & = f_{\eq,n} \bar{f}_{\eq,n} \Big(c_T m_n^2 + b_n \big(c_B(\up) {+} c_V^{\langle\mu\rangle} p_{\langle\mu\rangle}\big) \\
  & + c_E (u {\,\cdot\,}p)^2 + c_Q^{\langle\mu\rangle} (u {\,\cdot\,}p) p_{\langle\mu\rangle} + c_\pi^{\langle\munu\rangle} p_{\langle\mu} p_{\nu\rangle} \Big)\,,
\end{split}
\ee
where $p^{\langle\mu\rangle} = \Delta^\mu_\nu p^\nu$ and $p^{\langle\mu} p^{\nu\rangle} = \Delta^\munu_{\alpha\beta} p^\alpha p^\beta$, with $\Delta^\munu = g^\munu - u^\mu u^\nu$ and $\Delta^\munu_{\alpha\beta} =\frac{1}{2}\big(\Delta^\mu_\alpha \Delta^\nu_\beta{+}\Delta^\mu_\beta \Delta^\nu_\alpha \big) - \frac{1}{3}\Delta^\munu \Delta_{\alpha\beta}$, are purely spatial in the LRF and traceless. The irreducible coefficients in (\ref{eq:14_moment_irreducible}) are
\bs
\allowdisplaybreaks
\beal
  c_B &= u_\mu c^{\mu} \,,
\\
  c_V^{\langle\mu\rangle} &= \Delta^\mu_\nu c^{\nu} \,,
\\
  c_T &= g_\munu c^\munu \,,
\\
  c_E &= u_\mu u_\nu  c^{\munu} \,,
\\
  c_Q^{\langle\mu\rangle} &= u_\alpha \Delta^\mu_\beta c^{\alpha\beta} \,,
\\
  c_\pi^{\langle\munu\rangle} &= \Delta^\munu_{\alpha\beta} \, c^{\alpha\beta}\,.
\end{align}
\es
One solves for these coefficients as follows, by inserting Eq.~\eqref{eq:14_moment_irreducible} into the Landau matching conditions for the net baryon density $n_B$ and energy density $\ene$, the definition of the Landau frame $T^\munu u_\nu = \ene u^\mu$ (which implies a vanishing heat current $Q^\mu = 0$), and the kinetic definitions for the bulk viscous pressure $\Pi$, baryon diffusion current $V_B^\mu$ and shear stress tensor $\pi^\munu$:
\bs
\label{eq:matching}
\beal
  \delta n_B &= \sum_n \int_{p} b_n \, (\up) \delta f_n = 0 \,,
\\
  \delta \ene &= \sum_n \int_{p} (\up)^2 \delta f_n = 0 \,,
\\
  Q^\mu &= \sum_n \int_{p} (\up) \px \delta f_n = 0 \,,
\\
  \Pi &= \frac{1}{3} \sum_n \int_{p} (- p {\,\cdot\,} \Delta {\,\cdot\,} p) \delta f_n \,,
\\
  V_B^\mu &= \sum_n \int_{p} b_n \, \px \delta f_n \,,
\\
  \pi^\munu &= \sum_n \int_{p} p^{\langle\mu} p^{\nu\rangle} \delta f_n \,.
\end{align}
\es
Here the sum runs over the number of resonances, $n=1,\dots,N_R$, and $\int_{p} \equiv \int \dfrac{d^3p}{(2\pi\hbar)^3 E}$. After some algebra one obtains
\bs
\allowdisplaybreaks
\label{eq7}
\beal
  c_T &= \frac{\Pi \, \mathcal{P}}{\A_{21}\mathcal{P} + \N_{31}\mathcal{Q} + \J_{41}\mathcal{R}} \,,
\\
  c_B &= \frac{\Pi \, \mathcal{Q}}{\A_{21}\mathcal{P} + \N_{31}\mathcal{Q} + \J_{41}\mathcal{R}} \,,
\\
  c_E &= \frac{\Pi \, \mathcal{R} }{\A_{21}\mathcal{P} + \N_{31}\mathcal{Q} + \J_{41}\mathcal{R}} \,,
\\
  c_V^{\langle\mu\rangle} &= \frac{V_B^\mu \J_{41}}{\N^2_{31} - \M_{21} \J_{41}} \,,
\\
  c_Q^{\langle\mu\rangle} &= -\frac{V_B^\mu \N_{31}} {\N^2_{31} - \M_{21} \J_{41}} \,,
\\
  c_\pi^{\langle\munu\rangle} &= \frac{\pi^\munu}{2(\ene{+}\Peq)T^2}\,,
\end{align}
\es
where $\Peq(\ene,n_B)$ is the equilibrium pressure and
\bs
\allowdisplaybreaks
\label{eq8}
\beal
  \mathcal{P} = &\, \N^2_{30} - \J_{40}\M_{20} \,, 
\\
  \mathcal{Q} = &\, \B_{10}\J_{40} - \A_{20}\N_{30} \,, 
\\
  \mathcal{R} = &\, \A_{20}\M_{20} - \B_{10}\N_{30}\,. 
\end{align}
\es
The thermal integrals $\J_{kq}$, $\mathcal{N}_{kq}$, $\mathcal{M}_{kq}$, $\A_{kq}$, and $\B_{kq}$ are defined in Appendix~\ref{app:integrals}. It is well known that the coefficients of the 14--moment approximation are linearly proportional to the dissipative flows of the fluid~\cite{DeGroot:1980dk}. An obvious problem with truncating the $\delta f_n^{14}$ correction to first order is that it can overwhelm the local-equilibrium distribution at sufficiently high momentum, potentially turning the total distribution function negative. This problem grows worse with larger dissipative flows. In principle, one can systematically improve the moments expansion by including the non-hydrodynamic moments ($\int_p p^\mu p^\nu p^\lambda f_n$, etc.)~\cite{Denicol:2012cn}. However, using them as macroscopic input along with $J_B^\mu$ and $T^\munu$ would require solving a set of evolution equations for these higher-order moments together with the fluid dynamical simulation.

\subsection{First-order RTA Chapman--Enskog expansion}
\label{sec2b}

Another common approach to obtaining a linear $\delta f_n$ correction for the Cooper--Frye formula is the Chapman--Enskog expansion, which is a perturbative series of the form
\be
\label{eq:gradient_series}
  f_n =  f_{\eq,n} + \sum^\infty_{k=1} \delta f_n^{(k)} = f_{\eq,n} + \sum^\infty_{k=1} \epsilon^k h_n^{(k)}\,,
\ee 
where $\epsilon$ and $h_n^{(k)}$ are the expansion parameter and coefficients~\cite{Jaiswal:2014isa}. This series should satisfy the RTA Boltzmann equation~\cite{Anderson_Witting_1974}
\be
\label{eq:RTA}
  p^\mu \del_\mu f_n = - \frac{(\up)(f_n {-} f_{\eq,n})}{\tau_r}\,,
\ee
where we take the relaxation time $\tau_r(x)$ to be momentum and species independent. If the hydrodynamic gradients are small compared to the relaxation rate $\tau_r^{-1}$, the expansion parameter is the directional derivative
\be
\label{eq11}
  \epsilon = - s^\mu \del_\mu
\ee
with
\be
\label{eq:s}
  s^\mu(x) = \dfrac{\tau_r(x)}{u(x) {\,\cdot\,} p}\, p^\mu.
\ee
Inserting the ansatz \eqref{eq:gradient_series} into Eq.~\eqref{eq:RTA} one finds at first order in gradients
\be
\label{eq:df_RTA}
\delta f^{(1)}_n = - s^\mu \del_\mu f_{\eq,n} \,.
\ee 
Using Eq. (\ref{eq:feq}) and expanding the derivative one obtains
\be
\label{eq:df1}
\begin{split}
  \delta f_n^{(1)} &= - \tau_r f_{\eq,n} \bar{f}_{\eq,n} \bigg(b_n \dot{\alpha}_B + \frac{(\up)\,\dot{T}}{T^2} + \frac{(- p {\,\cdot\,} \Delta {\,\cdot\,} p)\theta}{3(\up)T} 
\\
  & + \frac{b_n \px \nabla_\mu \alpha_B}{\up} - \frac{\px (\dot{u}_\mu {-} \nabla_\mu {\ln} T)}{T} - \frac{\sigma_\munu p^{\langle\mu} p^{\nu\rangle}}{(\up)T}\bigg) \,,
\end{split}
\ee
where $\theta = \del_\mu u^\mu$ is the scalar expansion rate and $\sigma_\munu = \del_{\langle\mu} u_{\nu\rangle} \equiv \Delta^{\alpha\beta}_\munu \partial_\beta u_\alpha$ is the velocity shear tensor. We denote the LRF time derivative as $\dot{a} = u^\mu \del_\mu a$ and the spatial gradient in the LRF as $\nabla^\mu a = \Delta^\munu \del_\nu a$. To simplify Eq.~\eqref{eq:df1} in terms of the hydrodynamic quantities one makes use of the conservation equations for the net baryon number, energy and momentum,
\bs
\label{eq:conservation_laws}
\beal
  \partial_\mu J_B^\mu & = 0 \,,
\\
  u_\nu \partial_\mu T^\munu & = 0 \,,
\\
  \Delta^\mu_\nu \partial_\lambda T^{\lambda\nu} & = 0\,,
\end{align}
\es
to eliminate the time derivatives:
\bs
\label{eq15}
\beal
  \dot\alpha_B &\approx \mathcal{G} \theta\,,
\\
  \dot T &\approx \mathcal{F} \theta \,,
\\
  \dot u^\mu &\approx \nabla^\mu \ln{T} + \frac{n_B T}{\ene {+} \Peq} \nabla^\mu \alpha_B\,.
\end{align}
\es
Here we only keep the first-order terms; the coefficients $\mathcal{G}$ and $\mathcal{F}$ are listed in Appendix~\ref{app:conservation}. The spatial gradients are substituted by dissipative flows using the Navier--Stokes relations:
\bs
\allowdisplaybreaks
\label{eq16}
\beal
  \Pi &\approx - \zeta \theta \,,
\\
  V_B^\mu &\approx \kappa_B \nabla^\mu\alpha_B \,,
\\
  \pi^\munu &\approx 2 \eta \sigma^\munu \,.
\end{align}
\es
The first-order RTA Chapman--Enskog expansion \eqref{eq:df1} then reduces to 
\be
\label{eq:Chapman_Enskog}
\begin{split}
  \delta f^{\text{CE}}_n & =  f_{\eq,n} \bar{f}_{\eq,n}\bigg[\frac{\Pi}{\beta_\Pi}\bigg(b_n \mathcal{G} +\frac{(\up)\mathcal F}{T^2} + \frac{(-p \cdot \Delta \cdot p)}{3(\up)T}\bigg) 
\\
  & + \frac{V_B^\mu p_{\langle\mu\rangle}}{\beta_V}\bigg(\frac{n_B}{\ene {+} \Peq} - \frac{b_n}{\up}\bigg)+ \frac{\pi_\munu p^{\langle\mu} p^{\nu\rangle}}{2\beta_\pi (\up)T} \bigg]\,,
\end{split}
\ee
where $\beta_\Pi$, $\beta_V$ and $\beta_\pi$ are the ratios of the bulk viscosity, baryon diffusion coefficient and shear viscosity, respectively, to the relaxation time. It is straightforward to check that $\delta f^{\text{CE}}_n$ satisfies the Landau matching and frame conditions (\ref{eq:matching}a-c). The coefficients $\beta_\Pi$, $\beta_V$ and $\beta_\pi$ can be extracted by inserting Eq.~\eqref{eq:Chapman_Enskog} into Eqs.~(\ref{eq:matching}d-f). The resulting expressions are 
\bs
\allowdisplaybreaks
\label{eq18}
\beal
  \beta_\Pi &= \mathcal{G} n_B T + \frac{\mathcal{F} (\ene{+}\Peq)}{T} + \frac{5 \J_{32}}{3T} \,,
\\
  \beta_V &= \mathcal{M}_{11} - \frac{n_B^2 T}{\ene{+}\Peq} \,,
\\
  \beta_\pi &= \frac{\J_{32}}{T} \,.
\end{align}
\es
Like the 14--moment approximation, the first-order Chapman--Enskog expansion reproduces $J_B^\mu$ and $T^\munu$ exactly, but the total distribution function can turn negative at high momentum. In this case, one could improve the expansion scheme by adding higher-order gradient corrections, but this approach is tedious and will not be pursued here. 

\section{Modified equilibrium distribution} 
\label{sec3}
\subsection{Formulation}
\label{sec3a}

Both the 14--moment approximation (\ref{eq:14_moment_irreducible}) and the first-order RTA Chapman--Enskog expansion (\ref{eq:Chapman_Enskog}) solve the problem of parametrizing the momentum dependence of $\delta f_n(x,p)$ in terms of the available hydrodynamic information in such a way that the dissipative flows are exactly reproduced from the corresponding hydrodynamic moments of $\delta f_n$, but they also suffer from the negative probability problem at sufficiently high momenta. The latter is a direct consequence of truncating the expansion underlying each of the two approaches at first order in the dissipative flows. One way to circumvent this problem is to manipulate the argument of the exponential function in the local-equilibrium distribution \eqref{eq:feq} with viscous corrections, as was done in Ref.~\cite{Pratt:2010jt}. By construction, this avoids the negative probability problem but, as we will see, at the expense of not being able to match the dissipative flows exactly to the corresponding moments of this modified equilibrium distribution.

Our starting point is the RTA Chapman--Enskog expansion. Equation \eqref{eq:df_RTA} is the first-order term of the Taylor series
\be
\label{eq21}
  f_{\eq,n}(x{\,-\,}s, p) \approx f_{\eq,n}(x,p) - s^\mu \del_\mu f_{\eq,n}(x,p)\,,
\ee
with $s^\mu$ given by Eq.~(\ref{eq:s}). Therefore we have as a first approximation for the modified equilibrium distribution
\be
\label{eq:feqmod_start}
  f^\text{(mod)}_{\eq,n}(x,p) = f_{\eq,n}(x{\,-\,}s, p)\,,
\ee
which retains the same analytic form as $f_{\eq,n}$, except for a shift in the position arguments of the auxiliary fields $T(x)$, $\alpha_B(x)$ and $u^\mu(x)$. Note that the shift depends on the particle momentum $p^\mu$ at which the distribution function is evaluated. Assuming small gradients we can evaluate these corrections to first-order:
\bs
\allowdisplaybreaks
\label{eq:shifts}
\beal
  T(x{\,-\,}\ds) &\approx T(x) - \ds^\nu\del_\nu T(x) \,,
\\
  \alpha_B(x{\,-\,}\ds) &\approx \alpha_B(x) - \ds^\nu\del_\nu \alpha_B(x) \,,
\\
  u^\mu(x{\,-\,}\ds) &\approx u^\mu(x) - \ds^\nu\del_\nu u^\mu(x)\,.
\end{align}
\es
The perturbations can be rewritten as
\bs
\allowdisplaybreaks
\label{eq:perturbations}
\beal
  \delta T &= - \tau_r \dot T - \frac{\tau_r p^{\langle\mu\rangle}\nabla_\mu T}{\up} \,,
\\
  \delta \alpha_B &= - \tau_r \dot \alpha_B - \frac{\tau_r p^{\langle\mu\rangle}\nabla_\mu \alpha_B}{\up} \,,
\\\!\!\!\!
  \delta u {\,\cdot\,}p &= \frac{\tau_r \theta (- p {\,\cdot\,} \Delta {\,\cdot\,} p)}{3(\up)} - \tau_r \dot u_\mu  p^{\langle\mu\rangle} - \frac{\tau_r \sigma_\munu p^{\langle\mu} p^{\nu\rangle}}{\up}\,.
\end{align}
\es
As before, we can eliminate the time derivatives and spatial gradients by using the conservation laws and Navier--Stokes relations. To obtain more precise expressions for the perturbations, we first linearize the $\delta f_n$ correction in Eq.~\eqref{eq:feqmod_start} as
\be
\label{eq:feqmod_expand}
  \delta f_n \approx f_{\eq,n} \bar{f}_{\eq,n} \bigg(b_n \delta\alpha_B + \frac{(\up)\delta T}{T^2} - \frac{\delta u {\,\cdot\,}p}{T} \bigg)\,.
\ee
After substituting Eqs.~\eqref{eq:perturbations}, \eqref{eq15} and \eqref{eq16} in Eq.~\eqref{eq:feqmod_expand} we recover the RTA Chapman--Enskog expansion \eqref{eq:Chapman_Enskog}. We now compare Eqs.~\eqref{eq:Chapman_Enskog} and \eqref{eq:feqmod_expand} to read off the perturbations:\footnote{%
    We here drop the spatial temperature gradients arising from this procedure in Eqs.~(\ref{eq:perturbations}a,c) since they cancel in Eq.~\eqref{eq:feqmod_expand}.}
\bs
\label{eq:perturb}
\beal
  \delta T &= \frac{\Pi \mathcal{F}}{\beta_\Pi} \,,
\\
  \delta \alpha_B &= \frac{\Pi \mathcal{G}}{\beta_\Pi} - \frac{V_B^\mu p_{\langle\mu\rangle}}{\beta_V(\up)} \,,
\\\!\!\!\!\!\!\!\!
  \delta u {\,\cdot\,}p & ={-}\frac{\Pi({-}p{\cdot} \Delta{\cdot}p)}{3\beta_\Pi(\up)} - \frac{V_B^\mu p_{\langle\mu\rangle} n_B T}{\beta_V(\ene{+}\Peq)} - \frac{\pi_\munu p^{\langle\mu} p^{\nu\rangle}}{2\beta_\pi(\up)}.
\end{align}
\es
These corrections are now used to modify the exponent of Eq.~(\ref{eq:feqmod_start}) to obtain\footnote{%
    Equation~\eqref{eq:feqmod_1} is closely related to the ``maximum entropy" (ME) distribution recently proposed in Ref.~\cite{Everett:2021ulz}, which modifies the exponent of the local-equilibrium distribution with additional Lagrange multipliers. At first order in the dissipative flows, it was shown \cite{Everett:2021ulz} that the non-equilibrium corrections due to these Lagrange multipliers reduce to the perturbative terms in Eq.~\eqref{eq:feqmod_1}.}
\be
\label{eq:feqmod_1}
  f_{\eq,n}^{\text{(mod)}} = \frac{g_n}{\exp\Bigl[\dfrac{(u {+} \delta u) {\,\cdot\,} p}{T {+} \delta T} - b_n(\alpha_B {+} \delta\alpha_B)\Bigr] + \Theta_n}\,.
\ee
As long as the viscous corrections are not too large (say, $|\Pi| < 3\beta_\Pi$) this expression is positive definite, but it does not reproduce the components of $J_B^\mu$ and $T^\munu$ exactly since it contains higher-order terms beyond the first-order RTA Chapman--Enskog expansion. These discrepancies are at least second order in gradients and should therefore be negligible if the viscous corrections are small:
\bs
\allowdisplaybreaks
\label{eq:small_viscous_limit}
\beal
  |\Pi| & \ll \beta_\Pi \,,
\\
  \sqrt{V_{B,\mu} V_B^\mu} & \ll \beta_V \,,
\\
  \sqrt{\pi_\munu \pi^\munu} & \ll 2 \beta_\pi.
\end{align}
\es
However, Eq.~\eqref{eq:feqmod_1} turns out to quickly lose its usefulness for even moderate viscous corrections which are found to result in serious violations of the matching to $J_B^\mu$ and $T^\munu$. As it stands, Eq.~\eqref{eq:feqmod_1} is not yet sufficient and needs additional improvements in order to mitigate the errors arising from the higher-order terms.\footnote{%
    One option is to replace the perturbations in Eq.~\eqref{eq:feqmod_1} by the Lagrange multipliers in Ref.~\cite{Everett:2021ulz} and adjust them so that the maximum entropy distribution exactly reproduces the hydrodynamic moments. However, an exact (numerical) calculation of these Lagrange multipliers for moderately large viscous corrections is still outstanding.}

\subsection{Local momentum transformation and normalization}
\label{sec3b}

The main source of errors in the modified equilibrium distribution \eqref{eq:feqmod_1} comes from the momentum dependent terms in the perturbations  (\ref{eq:perturb}b,c). In the limit of small gradients, the deformations and shifts to the momentum space are linearly proportional to the dissipative flows. For larger viscous corrections, however, these effects propagate into the hydrodynamic quantities non-linearly. In order to control these errors, one should recast the momentum-dependent perturbations in such a way that the viscous corrections to the momentum space are strictly linear.

The local momentum transformation \eqref{eq:rescale}, first introduced in Ref.~\cite{Pratt:2010jt}, provides an effective way of dealing with the momentum-dependent perturbations in Eq.~\eqref{eq:perturb}. Following their prescription, we rewrite Eq.~\eqref{eq:feqmod_1} as
\be
\label{eq:feqmod_2}
\!\!\!\!
  f_{\eq,n}^{\text{PTM}} = \frac{\mathcal{Z}_n g_n}{\exp\biggl[\dfrac{\sqrt{\bm{p}'^2 {+} m_n^2}}{T {+} \beta_\Pi^{-1}\Pi \mathcal{F}} - b_n \Big(\alpha_B {+} \dfrac{\Pi \mathcal{G}}{\beta_\Pi}\Big)\biggr] + \Theta_n}.\!
\ee
Starting from a \textit{thermal} distribution of momenta $p'$ we construct a map $p = M(p^{\,\prime})$ to the LRF momenta $p$ such that the deformations and shifts are linearly proportional to the dissipative flows. Under the constraint of reducing to the momentum-dependent perturbations in the limit of small gradients, one finds the following transformation between $p$ and $p'$:
\be
\label{eq:general_rescaling}
  p_i = A_{ij} p'_j - q_i \sqrt{\bm{p}'^2 {+} m_n^2} + b_n T a_i \,,
\ee
where
\bs
\allowdisplaybreaks
\label{eq:rescale_matrix}
\beal
  A_{ij} &= \biggl(1{+}\frac{\Pi}{3\beta_\Pi}\biggr)\,\delta _{ij} + \frac{\pi_{ij}}{2\beta_\pi} \,,
\\
  q_{i} &= \frac{ V_{B,i} \, n_B T}{\beta_V(\ene{+}\Peq)} \,,
\\
a_{i} &= \frac{V_{B,i}}{\beta_V}\,,
\end{align}
\es
with $V_{B,i} =- X_i {\,\cdot\,} V_B$ being the LRF spatial components of the baryon diffusion current. The deformation matrix $A_{ij}$ is identical to the one in Eqs.~(\ref{eq:rescale},\ref{eq:Aij}), i.e. $\lambda_\Pi=1/(3\beta_\Pi)$ and $\lambda_\pi=1/(2\beta_\pi)$. Compared to Eqs.~(\ref{eq:rescale},\ref{eq:Aij})~\cite{Pratt:2010jt}, Eqs.~(\ref{eq:general_rescaling},\ref{eq:rescale_matrix}) are further generalized to include baryon diffusion effects. 

For the normalization factor $\mathcal{Z}_n$ we fix the particle density of the modified equilibrium distribution to that given by the RTA Chapman--Enskog expansion,
\be
\label{eq33}
  n^{(1)}_n  = n_{\eq,n} + \frac{\Pi}{\beta_\Pi}\bigg(n_{\eq,n}+ \N_{10,n} \mathcal G + \frac{\J_{20,n} \mathcal F}{T^2} \bigg)\,,
\ee
which contains a bulk viscous correction to the equilibrium particle density $n_{\eq,n}(T,\alpha_B)$~\cite{Pratt:2010jt, Bernhard:2018hnz}. This suppresses the nonlinear shear and bulk viscous corrections to the modified particle density,
\be
\label{eq:mod_density}
  n^{\text{PTM}}_{n} = \mathcal{Z}_n \, {\det}A \,\cdot\, n_{\eq,n}(T{+}\beta_\Pi^{-1}\Pi \, \mathcal{F},\alpha_B{+}\beta_\Pi^{-1}\Pi \, \mathcal{G}),
\ee
which (if $\mathcal{Z}_n = 1$) are another major source of errors in matching to the hydrodynamic output. Our choice for the normalization factor is therefore
\be
\label{eq:renorm}
  \mathcal{Z}_n = \frac{1}{{\det}A} \times \frac{n_n^{(1)}}{n_{\eq,n}(T{+}\beta_\Pi^{-1}\Pi \, \mathcal{F},\alpha_B{+}\beta_\Pi^{-1}\Pi \, \mathcal{G})}\,,
\ee
which is similar to Eq.~\eqref{eq:Z_PT} except it contains an additional factor. One should keep in mind that the linearized density $n^{(1)}_n$ of light hadrons, especially pions, may turn negative if the bulk viscous pressure is too negative.

The ``PTM distribution" (\ref{eq:feqmod_2}) cannot be applied in every situation. The requirements for using $f_{\eq,n}^\text{PTM}$ in the Cooper--Frye formula are that the determinant of the Jacobian
\be
\label{eq:Jacobian}
  \det\bigg(\frac{\partial p_i}{\partial p'_j}\bigg) 
  = {\det}A\,\bigg(1{\,-\,}\frac{q_i A^{-1}_{ij}p'_j}{\sqrt{\bm{p}'^2 {+} m_n^2}}\bigg)
\ee
is positive for any value of $p'$, the deformation matrix $A_{ij}$ is invertible, and the normalization factor $\mathcal{Z}_n$ is non-negative. These conditions can be violated when the dissipative flows are too large. What to do in this case will be discussed in Sec.~\ref{sec5} where we study certain situations where the modified equilibrium distribution breaks down.

\subsection{Reproducing the hydrodynamic quantities}
\label{sec3c}

As already noted, the nonlinear dependence of the modified equilibrium distribution (\ref{eq:feqmod_2}) on the dissipative flows makes it impossible to achieve an exact matching of the hydrodynamic moments of this distribution with all components of $T^\munu$ and $J_B^\mu$. This is different from the linearized parametrizations described in Sec.~\ref{sec2}. In this subsection, we study the matching violations for hadrons emitted from a stationary time-like freeze-out cell with the modified equilibrium distribution \eqref{eq:feqmod_2}.

\begin{figure*}[t]
\centering
\includegraphics[width=\textwidth]{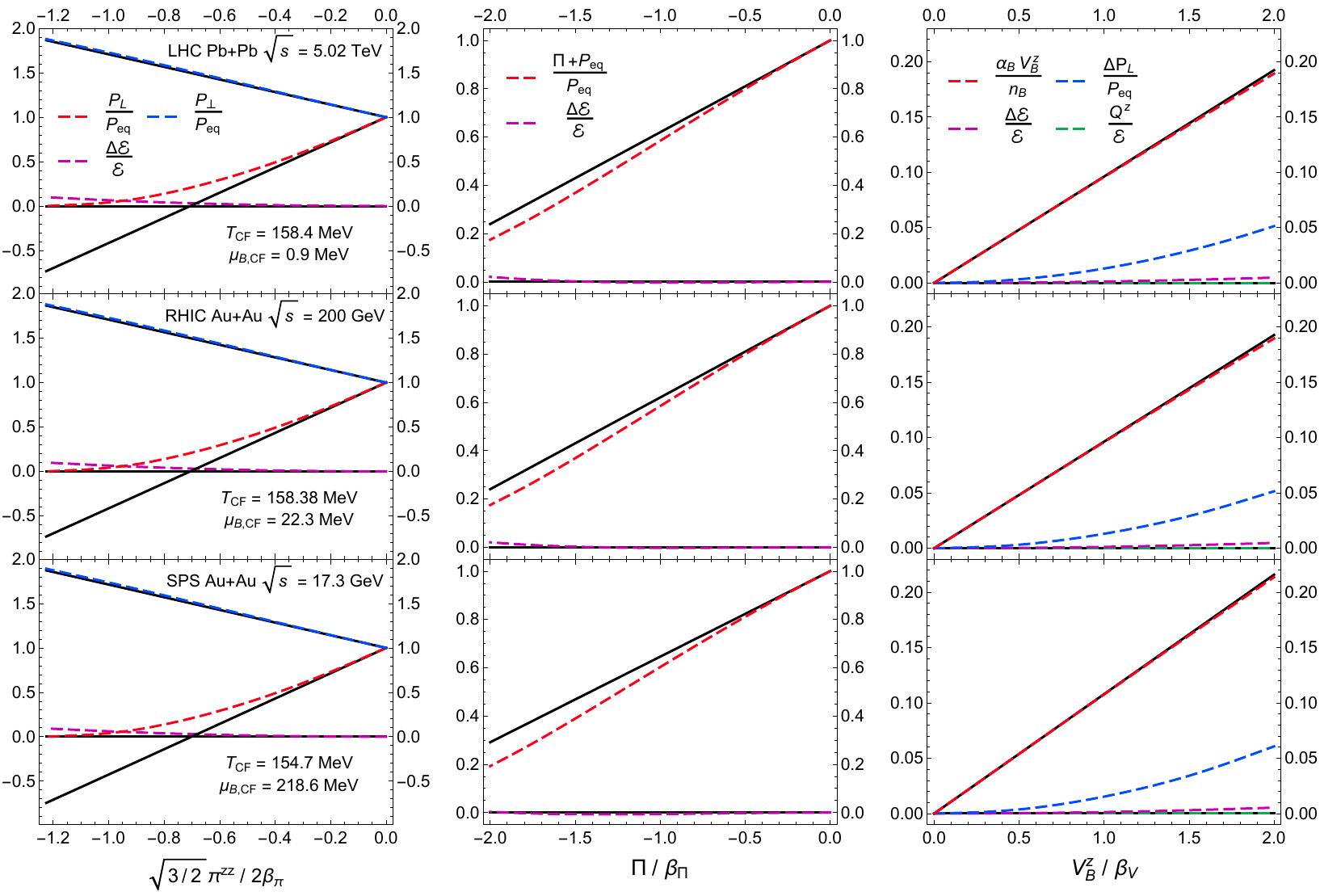}
\caption{
    The reproduction of selected components of the net baryon current $J_B^\mu$ and energy-momentum tensor $T^\munu$ using the Cooper--Frye prescription with the PTM distribution (\ref{eq:feqmod_2}), for a stationary hadron resonance gas subject to either shear stress (left column), bulk viscous pressure (middle column) or a baryon diffusion current (right column). The temperature and baryon chemical potential $(T_\text{CF},\mu_{B,\text{CF}})$ at chemical freeze-out are indicated in the legend (top to bottom rows). Solid black curves represent the hydrodynamic input (i.e. the target $J_B^\mu$ and $T^\munu$) while the dashed color lines are the output components of $J_B^\mu$ and $T^\munu$ obtained by computing the corresponding moments of $f_{\eq}^{\text{PTM}}$. 
\label{hydro_output}
}
\end{figure*}

The system is composed of the set of hadron species included in the hadronic afterburner code URQMD~\cite{Bleicher:1999xi} ($N_R \sim 320$). These hadrons are assumed to be produced from a fluid in chemical equilibrium at the chemical freeze-out (CF) temperature $T_\text{CF}$ and baryon chemical potential $\mu_{B,\text{CF}}$, which can vary across different collision systems. The values used here (shown in the legend of Figure~\ref{hydro_output}) are taken from a statistical model fit to the hadron abundance ratios measured in Pb+Pb collisions at LHC energies ($\sqrt{s_\mathrm{NN}}= 5.02$ TeV) and in Au+Au collisions at RHIC and SPS energies ($\sqrt{s_\text{NN}} = 200$ GeV and 17.3 GeV, respectively)~\cite{Andronic:2017pug}. For clarity, we vary either the input shear stress, bulk viscous pressure or baryon diffusion current while fixing the other dissipative flows to zero. 

In the first example (left column of Fig.~\ref{hydro_output}) we consider the case where in a given freeze-out cell the hydrodynamic $T^\munu$ features a negative pressure anisotropy $\frac{2}{3}(\PL{-}\Pperp) = \pi^{zz} = -2\pi^{xx} = -2\pi^{yy}$, combined with zero bulk viscous pressure and baryon diffusion current. Starting from equilibrium at the right edge of the plots, we decrease $\pi^{zz}$ going left, causing the longitudinal pressure $\PL = \Peq + \pi^{zz}$ to decrease and the transverse pressure $\Pperp = \Peq + \frac{1}{2}(\pi^{xx} {+} \pi^{yy})$ to increase. The energy density should not change due to the Landau matching condition (\ref{eq:matching}a). These are the hydrodynamic components of $T^\munu$ that we aim to reproduce, shown by the solid black curves. The colored dashed curves are the kinetic outputs, calculated as moments of the modified equilibrium distribution with these hydrodynamic inputs.\footnote{%
    Due to the normalization factor~\eqref{eq:renorm} the modified equilibrium distribution conserves the net baryon number exactly ($\delta n_B{\,=\,}0$).}$^,$\footnote{%
    For only shear stress inputs, the outputs $Q^\mu$ and $V_B^\mu$ vanish by symmetry and are therefore not of interest here.} 
One sees that for small $\pi^{zz}$ the kinetic output closely follows the initial hydrodynamic input. This is because in the limit of small dissipative flows the modified equilibrium distribution reduces to the linear RTA Chapman--Enskog expansion for which the kinetic output reproduces the hydrodynamic input exactly. As $\pi^{zz}$ further decreases to larger negative values, the kinetic output $T^\munu$ begins to deviate from the hydrodynamic target. In particular, for a positive definite distribution function such as Eq.~(\ref{eq:feqmod_2}) the kinetic output $\PL$ stays always above zero even if the hydrodynamic input for $\pi^{zz}$ is so large and negative that the total longitudinal pressure in the fluid is negative. On the other hand, the kinetic outputs for $\ene$ and $\Pperp$ agree well with their hydrodynamic targets even for large pressure anisotropies. These trends are confirmed for all three combinations $(T_\text{CF},\mu_{B,\text{CF}})$ studied in the three rows of Fig.~\ref{hydro_output}. Technically, the modified equilibrium distribution breaks down completely for $\pi^{zz} \leq -2\beta_\pi$ (which causes ${\det}A \leq 0$), and one should not expect $f_{\eq,n}^{\text{PTM}}$ to work well for very large pressure anisotropies. However, for moderately large $|\pi^{zz}| \leq \frac{1}{2}\Peq$, $f_{\eq,n}^{\text{PTM}}$ reproduces the target $T^\munu$ components quite well, with $|\Delta\PL| / \PL  \leq 9.0\%$, $|\Delta\Pperp| / \Pperp \leq 0.5\%$ and $|\Delta\ene| / \ene \leq 0.8\%$.\footnote{%
    We also successfully replicated the shear stress test shown in Fig.~2 of Ref.~\cite{Pratt:2010jt}.}

%
%

In the middle column of Fig.~\ref{hydro_output} we vary the bulk viscous pressure $\Pi$ at vanishing shear stress and baryon diffusion. We checked that for this hydrodynamic input the kinetic outputs for $\pi^\munu$, $Q^\mu$ and $V_B^\mu$ vanish. The plots show that, for all three choices of freeze-out parameters, the energy matching condition $\Delta\ene = 0$ holds very well even for large (negative) values of $\Pi$. The kinetic output for the total pressure $\Peq + \Pi$ tends to somewhat underpredict the hydrodynamic target, the error staying below 10\% for moderately large bulk viscous pressures $|\Pi| \leq \frac{1}{2}\Peq$. For more negative input values of $\Pi$, $\Pi \lesssim -2\beta_\Pi$, the particle density for pions turns negative and the modified equilibrium distribution becomes invalid.

In the right column of Fig.~\ref{hydro_output} we test the reproduction of the baryon diffusion current (taken arbitrarily to point in the $z$--direction) at zero shear and bulk viscous stress. One sees that the hydrodynamic input for the ratio $\alpha_B V_B^z /n_B$ (which is approximately the same for all three freeze-out parameter pairs) is well reproduced by the kinetic output from the  modified equilibrium distribution. The energy matching and Landau frame ($Q^z=0$) conditions are also reproduced with excellent precision. However, a nonzero input for $V_B^z \ne 0$ also generates nonzero kinetic outputs for the bulk and shear viscous stresses even when their hydrodynamic inputs are zero: For $V_B^z/\beta_V=2$ we find a 5\% positive difference $\Delta \PL/\Peq$ between the hydrodynamic input and the kinetic output for the longitudinal pressure of which, after decomposition into shear and bulk viscous contributions, 1.5\% can be attributed to an induced bulk viscous pressure $\Pi$ and 3.5\% to an induced shear stress $\pi^{zz}$, both with positive signs. For moderately large baryon diffusion currents $|V_B^z| \leq \beta_V$, the hydrodynamic output errors are  $|\Delta V_B^z / V_B^z| \leq 0.5\%$, $|\Delta\PL| / \Peq \leq 1.3\%$, $|\Delta\ene| / \ene \leq 0.2\%$ and $|Q^z| / \ene \approx 0$.


%
\begin{figure}[t]
\includegraphics[width=1.0\linewidth]{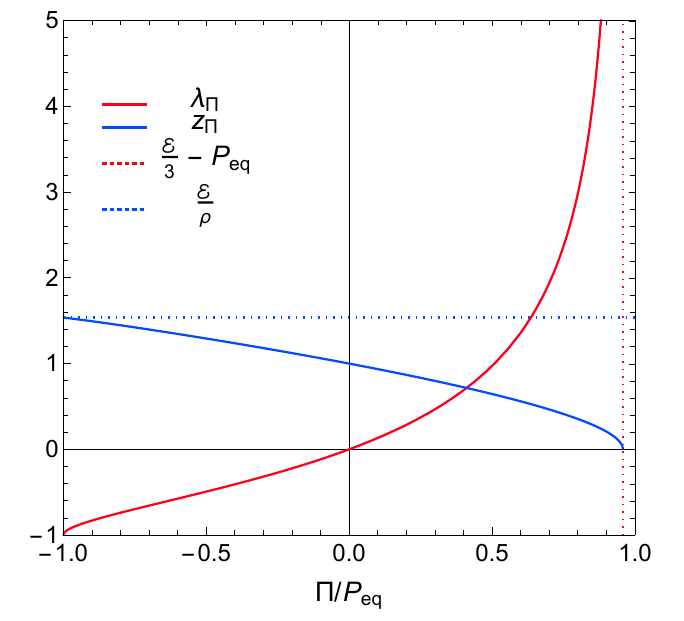}
\caption{
    The isotropic scale parameter $\lambda_\Pi$ (solid red) and normalization factor $z_\Pi$ (solid blue) as a function of $\Pi/\Peq$ are computed for a hadron resonance gas at a fixed temperature $T = 150$ MeV. The dotted blue and red lines are the upper bounds for $z_\Pi$ and $\Pi / \Peq$, respectively. This plot corresponds to Fig.~3.12 in Ref.~\cite{Bernhard:2018hnz}, where $\Delta \langle p \rangle / \langle p \rangle_0 = \lambda_\Pi$ and $\Delta n/n_0 = z_\Pi - 1$. 
\label{FJonah}
}
\end{figure}
%

\subsection{Pratt--Torrieri--Bernhard distribution}
\label{sec3d}

Another variant of Pratt and Torrieri's idea \cite{Pratt:2010jt} was implemented by Bernhard in Ref.~\cite{Bernhard:2018hnz}:
\be
\label{eq:Jonah}
  f_{\eq,n}^{\text{PTB}} = \frac{\mathcal{Z} g_n}{\exp\Bigl[\dfrac{\sqrt{\bm{p}'^2 {+} m_n^2}}{T}\Bigr] + \Theta_n}\,.
\ee
In this ``PTB distribution" the baryon chemical potential and diffusion current are neglected ($\alpha_B = V_B^\mu = 0$), and the effective temperature is not modified ($\delta T = 0$). The momentum transformation rule is $p_i = A_{ij} p^\prime_j$ with
\be
\label{eq:lambda}
  A_{ij} = (1 {+} \lambda_\Pi)\delta_{ij} + \frac{\pi_{ij}}{2\beta_\pi} \,.
\ee
The shear stress modification is the same as in Eq.~(\ref{eq:rescale_matrix}a) while the bulk pressure term is replaced by the isotropic scale parameter $\lambda_\Pi$. The normalization factor $\mathcal{Z}$ is taken as species-independent:
\be
\mathcal{Z} = \frac{z_\Pi}{{\det}A}\,,
\ee
leaving particle abundance ratios unchanged from their chemical equilibrium values. The parameters $\lambda_\Pi$ and $z_\Pi$ are fixed such that in the absence of shear stress ($\pi_{ij} = 0$) the energy density and bulk viscous pressure are exactly matched~\cite{Bernhard:2018hnz}:
\bs
\label{eq:bulkparameters}
\beal
  \ene'(\lambda_\Pi, z_\Pi,T) &= \ene\,,  
\\
  \mathcal{P}'(\lambda_\Pi, z_\Pi,T) &= \Peq + \Pi\,;
\end{align}
\es
here $\ene'$ and $\mathcal{P}'$ are the kinetic theory output for the energy density and isotropic pressure computed from the PTB distribution \eqref{eq:Jonah}. 

The parametrization procedure is as follows: For the distribution function \eqref{eq:Jonah} to be well-defined the isotropic scale parameter must satisfy $\lambda_\Pi \in (-1,\infty)$. At a given temperature $T$ one sets up a grid in $\lambda_\Pi$ and computes the corresponding values for $z_\Pi$ and $\Pi$ numerically by rewriting Eqs.~(\ref{eq:bulkparameters}a,b) as
\bs
\allowdisplaybreaks
\label{eq:zbulk}
\beal
  z_\Pi &= \frac{\ene}{\mathcal{L}_{20}(\lambda_\Pi, T)} \,, 
\\
  \Pi &= z_\Pi \mathcal{L}_{21}(\lambda_\Pi, T) - \Peq \,.
\end{align}
\es
The functions $\mathcal{L}_{kq}$ are defined in Appendix~\ref{app:integrals}. One can then interpolate the data with respect to $\Pi$ to construct the functions $\lambda_\Pi(\Pi; T)$ and $z_\Pi(\Pi; T)$; an example is shown in Figure~\ref{FJonah}. Generally, these are nonlinear functions of $\Pi$. In the limit of small bulk viscous pressure, they linearize to
\bs
\allowdisplaybreaks
\label{dz}
\beal
  z_\Pi &\approx  1 + \delta z_\Pi = 1 - \frac{3\Pi \Peq}{5 \beta_\pi \ene - 3\Peq(\ene {+} \Peq)} \,,
\\
  \lambda_\Pi &\approx \delta \lambda_\Pi = \frac{\Pi \ene}{5 \beta_\pi \ene - 3\Peq(\ene {+} \Peq)}\,.
\end{align}
\es
Equations~(\ref{eq:zbulk}a,b) imply that $z_\Pi$ and $\Pi$ are bounded by
\bs
\allowdisplaybreaks
\label{eq:bounds}
\beal
  0 < &\,z_\Pi < \frac{\ene}{\rho}\, ,
\\
  -\Peq < &\,\Pi < \frac{\ene}{3} - \Peq \,,
\end{align}
\es
where $\rho = \sum_n^{N_R} m_n n_{\eq,n}$ is the equilibrium mass density. When $\Pi$ lies outside the bound\footnote{%
    Violations of the upper bound in Eq.~(\ref{eq:bounds}b) can occur at the transition from a conformal pre-hydrodynamic model to non-conformal viscous hydrodynamics, where the mismatch between the conformal and QCD equations of state gives exactly $\Pi = \ene/3 - \Peq(\ene)$.}
(\ref{eq:bounds}b) or ${\det}A \leq 0$, we consider the modified equilibrium distribution \eqref{eq:Jonah} to have broken down. As will be discussed in Sec.~\ref{sec5b},  we then resort to linearizing Eq.~\eqref{eq:Jonah} around local equilibrium, i.e. we write $f_{\eq,n}^{\text{PTB}}\approx f_{\eq,n}+\delta f_n$, where
\be
\label{eq:linear_Jonah}
\begin{split}
  \delta f_n &= f_{\eq,n} \left(\delta z_\Pi - 3 \delta \lambda_\Pi\right) + 
\\
  & \, f_{\eq,n} \bar{f}_{\eq,n}\left(\frac{\delta \lambda_\Pi (-p \cdot \Delta \cdot p)}{(u\cdot p)T} + \frac{\pi_\munu p^{\langle\mu} p^{\nu\rangle}}{2\beta_\pi (u\cdot p)T}\right)\,.
\end{split}
\ee

The effectiveness of the PTB distribution (\ref{eq:Jonah}) in reproducing the components of a given energy-momentum tensor $T^\munu$ was studied in Ref.~\cite{Bernhard:2018hnz} to which we refer the reader for comparison with Fig.~\ref{hydro_output}.

\section{Modified anisotropic distribution}
\label{sec4}
\subsection{Anisotropic hydrodynamics}
\label{sec4a}

In this section we apply the formulation developed in Sec.~\ref{sec2} to modify an anisotropic distribution function that has already been deformed at leading order. In anisotropic hydrodynamics, the distribution function is decomposed into a momentum-anisotropic leading order term and a residual correction~\cite{Bazow:2013ifa, Molnar:2016vvu, McNelis:2018jho} as follows:
\be
\label{eq:vah}
  f_n(x,p) = f_{a,n}(x,p) + \delta\tilde{f}_n(x,p)\,.
\ee
Here the {\it anisotropic} leading order distribution $f_{a,n}$ is taken as\footnote{%
    For simplicity, we set in this Section the effective baryon chemical potential to $\tilde{\mu}_B = 0$.}
\be
\label{eq:fa}
  f_{a,n}(x,p) = \frac{g_n}{\exp\biggl[\dfrac{\sqrt{p \cdot \Omega(x) \cdot p + m_n^2}}{\Lambda(x)}\biggr] + \Theta_n}\,.
\ee
Similar to the modified equilibrium distribution \eqref{eq:feqmod_2}, it has an effective temperature $\Lambda$ and a momentum transformation encoded in the ellipsoidal tensor
\be
\Omega_\munu = \frac{Z_\mu Z_\nu}{\alpha_L^2} - \frac{\Xi_\munu}{\alpha_\perp^2}\,.
\ee
Here $Z^\mu{\,\equiv\,}X_3^\mu$ is the longitudinal basis vector and $\Xi^\munu = g^\munu{-}u^\mu u^\nu{+}Z^\mu Z^\nu$ the associated transverse spatial projector~\cite{Molnar:2016vvu}. The momentum anisotropy parameters $\alpha_L$ and $\alpha_\perp$ deform the longitudinal and transverse momentum space, respectively. The anisotropic distribution \eqref{eq:fa} can be rewritten more conveniently as
\be
\label{eq:fa2}
  f_{a,n} = \frac{g_n}{\exp\Bigl[\dfrac{\sqrt{\bm{p}'^2 {+} m_n^2}}{\Lambda}\Bigr] + \Theta_n} \,,
\ee
where $p_i = A_{ij} p'_j$ with
\be
\label{eq:Aij_mod}
   A_{ij} = \alpha_\perp \delta_{ij} + (\alpha_L {-} \alpha_\perp) Z_i Z_j \,,
\ee
and $Z_i = - X_i {\,\cdot\,} Z = (0,0,1)$ being the LRF components of the longitudinal basis vector. Together, the anisotropic parameters $\Lambda$, $\alpha_L$ and $\alpha_\perp$ are adjusted such that $f_{a,n}$ completely captures the energy density $\ene$ and the longitudinal and transverse pressures, $\PL$ and $\Pperp$, in the energy-momentum tensor~\cite{Molnar:2016vvu,McNelis:2018jho}
\be
  \!\!\!\!
  T^\munu = \ene u^\mu u^\nu {+} \PL Z^\mu Z^\nu {-} \Pperp \Xi^\munu + 2\Wperp^{(\mu} Z^{\nu)} + \piperp^\munu.\!
\ee
The residual correction $\delta \tilde{f}_n$ does not contribute to $\ene$, $\PL$ and $\Pperp$ but accounts for the longitudinal momentum diffusion current $\Wperp^\mu$ and the transverse shear stress tensor $\piperp^\munu$:
\bs
\allowdisplaybreaks
\beal
  \Wperp^\mu &= - \Xi^\mu_\alpha Z_\nu T^{\alpha\nu} \,,
\\
  \piperp^\munu &= \Xi^\munu_{\alpha\beta} T^{\alpha\beta}\,.
\end{align}
\es
Here $\Xi^\munu_{\alpha\beta} = \frac{1}{2}\big(\Xi^\mu_\alpha \Xi^\nu_\beta + \Xi^\nu_\beta \Xi^\mu_\alpha - \Xi^\munu \Xi_{\alpha\beta}\big)$ is the double-transverse traceless  projector. An expression for $\delta \tilde{f}_n$ can be derived using a linearization approach, such as the Chapman--Enskog expansion. After substituting Eq.~\eqref{eq:vah} in the RTA Boltzmann equation~\eqref{eq:RTA}, the first-order expression for $\delta \tilde{f}_n$ is
\be
\label{eq:ACE1}
  \delta\tilde{f}_n = - s^\mu \del_\mu f_{a,n} - (f_{a,n} {-} f_{\eq,n})\,,
\ee
where the second-order term $-s^\mu \del_\mu \delta \tilde{f}_n$ was neglected. The first term in Eq.~\eqref{eq:ACE1} can be expanded into
\be
  - s^\mu \del_\mu f_{a,n} = f_{a,n} \bar{f}_{a,n} \left(\dfrac{E_{a,n} \delta\Lambda}{\Lambda^2} - \frac{p \cdot\delta\Omega\cdot p}{2 E_{a,n} \Lambda}\right)\,,
\ee
where $\bar{f}_{a,n} {\,=\,} 1 {\,-\,} g_n^{-1} \Theta_n f_{a,n}$, $E_{a,n} {\,=\,} \sqrt{p {\,\cdot\,} \Omega {\,\cdot\,} p {+} m_n^2}$,
\begin{eqnarray}
  \frac{p \cdot \delta \Omega \cdot p}{2}  &=& \frac{(\alpha_L^2 {-} \alpha_\perp^2)(-Z\cdot p)(\delta Z \cdot p)}{\alpha_\perp^2 \alpha_L^2} + \frac{(u {\,\cdot\,}p)(\delta u {\,\cdot\,}p)}{\alpha_\perp^2} 
\nonumber\\
  && - \frac{\delta \alpha_L (-Z \cdot p)^2}{\alpha_L^3} - \frac{\delta \alpha_\perp (-p \cdot \Xi \cdot p)}{\alpha_\perp^3} \,,
\end{eqnarray}
and the perturbations $\delta u^\mu$ and $\delta Z^\mu$ are\footnote{%
    We find that the $\delta\tilde{f}_n$ terms $\propto$ ($\delta \Lambda$, $\delta \alpha_L$, $\delta \alpha_\perp$) do not contribute to $\Wperp^\mu$ and $\piperp^\munu$ so we can effectively set these perturbations to zero.}
\bs
\allowdisplaybreaks
\label{eq:dUdZ_full}
\beal
  \delta u {\,\cdot\,}p &= \frac{-\tau_r}{\up}\Big[(u\cdot p)\big((-Z\cdot p)Z^\mu {+} p^{\{\mu\}}\big)\dot{u}_\mu 
\\ \nonumber
  & - (-Z\cdot p)^2 \theta_L - \frac{1}{2}(-p\cdot\Xi\cdot p)\theta_\perp 
\\ \nonumber
  & + p^{\{\mu} p^{\nu\}} \sigma_{\perp,\munu} + (-Z\cdot p)p_{\{\mu\}}\big(Z_\nu \nabla_\perp^\mu u^\nu {-} D_z u^\mu\big) \Big] \,,
\\ 
  \delta Z \cdot p &= \frac{-\tau_r}{\up} \Big[{-}(u\cdot p)^2 Z_\mu \dot{u}^\mu + (u {\,\cdot\,}p)(-Z\cdot p)\theta_L
\\ \nonumber
  &+ (u {\,\cdot\,}p)(-Z\cdot p)\theta_L  
  + p^{\{\mu} p^{\nu\}} \sigma_{z,\munu}
\\ \nonumber
  &-(-Z\cdot p)p_{\{\mu\}}D_z Z^\mu - \frac{1}{2}(-p\cdot\Xi\cdot p)\nabla_{\perp\mu}Z^\mu 
\\ \nonumber
  &+ (u\cdot p)p_{\{\mu\}}\big(\dot{Z}^\mu {-} Z_\nu \nabla_\perp^\mu u^\nu\big) 
   \Big]\,,
\end{align}
\es
where $p^{\{\mu\}} \equiv \Xi^\mu_\nu\, p^\nu$, $p^{\{\mu} p^{\nu\}} \equiv \Xi^\munu_{\alpha\beta}\, p^\alpha p^\beta$, $\theta_L = Z_\mu D_z u^\mu$ is the longitudinal expansion rate, $\theta_\perp = \nabla_{\perp,\mu} u^\mu$ is the transverse expansion rate, $D_z = - Z {\,\cdot\,} \del$ is the LRF longitudinal derivative, $\nabla_\perp^\mu = \Xi^\munu \del_\nu$ is the transverse spatial gradient, $\sigma_{\perp,\munu} = \Xi^{\alpha\beta}_\munu \del_\alpha u_\beta$ is the transverse velocity shear tensor and $\sigma_{z,\munu} = \Xi^{\alpha\beta}_\munu \del_\alpha Z_\beta$~\cite{Molnar:2016vvu}. For the sake of simplicity, we only consider the terms with nonzero contributions to $\Wperp^\mu$ and $\piperp^\munu$, whose kinetic theory definitions are
\bs
\label{eq:Wpi_defs}
\beal
  \Wperp^\mu &= \sum_n \int_p (-Z\cdot p)p^{\{\mu\}} \delta\tilde{f}_n \,,
\\
  \piperp^\munu &= \sum_n \int_p p^{\{\mu} p^{\nu\}} \delta\tilde{f}_n\,.
\end{align}
\es
By symmetry, the term $(f_{a,n} {-} f_{\eq,n})$ in Eq.~\eqref{eq:ACE1} and corrections $\propto (\delta\Lambda, \delta\alpha_L, \delta\alpha_\perp)$ have zero contributions, so we can effectively eliminate them. The only nonzero contributions from $\delta u {\,\cdot\,}p$ and $\delta Z {\,\cdot\,} p$ in Eqs.~(\ref{eq:dUdZ_full}a,b) are 
\bs
\allowdisplaybreaks
\label{eq:dUdZ}
\beal
  \delta u {\,\cdot\,}p \rightarrow&  - \frac{\tau_r(-Z \cdot p)p_{\{\mu\}}\left(Z_\nu \nabla_\perp^\mu u^\nu {-} D_z u^\mu \right)}{\up} 
\\ \nonumber
  &- \frac{\tau_r p_{\{\mu} p_{\nu\}}\sigma_\perp^\munu}{\up} \,,
\\
  \delta Z \cdot p \rightarrow& - \tau_r p_{\{\mu\}}\big(\dot{Z}^\mu {-} Z_\nu \nabla_\perp^\mu u^\nu\big)\,.
\end{align}
\es
The $\delta \tilde{f}_n$ correction then reduces to
\be
\label{eq:ACEgrad}
\begin{split}
  \delta\tilde{f}_n &= \frac{\tau_r f_{a,n}\bar{f}_{a,n}}{E_{a,n}\Lambda}\left[\frac{p_{\{\mu} p_{\nu\}}\sigma_\perp^\munu}{\alpha_\perp^2} {-} \frac{({-}Z{\,\cdot\,} P)p_{\{\mu\}} \kappa_\perp^\mu}{\alpha_\perp \alpha_L}\right]\,,
\end{split}
\ee
where
\be
  \kappa_\perp^\mu = \frac{(\alpha_\perp^2 {-} \alpha_L^2)\Xi^\mu_\nu \dot{Z}^\nu {-} \alpha_\perp^2 Z_\nu \nabla_\perp^\mu u^\nu {+} \alpha_L^2 \Xi^\mu_\nu D_z u^\nu}{\alpha_\perp \alpha_L}\,.
\ee
The gradients $\sigma_\perp^\munu$ and $\kappa_\perp^\mu$ are proportional to the ``Navier--Stokes" values for $\piperp^\munu$ and $\Wperp^\mu$, respectively. After inserting Eq.~\eqref{eq:ACEgrad} in Eq.~\eqref{eq:Wpi_defs} one obtains
\be
  \piperp^\munu =  2 \tau_r \beta_\pi^\perp \sigma_\perp^\munu \,,\qquad
  \Wperp^\mu =  \tau_r \beta_W^\perp \kappa_\perp^\mu\,,
\ee
where
\be
  \beta_\pi^\perp = \frac{\mathcal{J}_{402-1}}{\alpha_\perp^2\Lambda} \,,
  \qquad
  \beta_W^\perp = \frac{\mathcal{J}_{421-1}}{\alpha_L \alpha_\perp \Lambda},
\ee
and the integrals $\mathcal{J}_{krqs}$ are defined in Appendix~\ref{app:integrals}. Thus, the first-order anisotropic RTA Chapman--Enskog expansion is
\be
\label{eq:ACE2}
  \!\!\!\!
  \delta\tilde{f}_n = f_{a,n} \bar{f}_{a,n}\left(\frac{p_{\{\mu} p_{\nu\}} \piperp^\munu}{2 \beta_\pi^\perp E_{a,n} \alpha_\perp^2 \Lambda} {-} \frac{(-Z \cdot p)p_{\{\mu\}} \Wperp^\mu}{\beta_W^\perp E_{a,n} \alpha_\perp \alpha_L \Lambda}\right)\!.\!\!
\ee
One can show that Eq.~\eqref{eq:ACE2} satisfies the anisotropic matching conditions for $\ene$, $\PL$ and $\Pperp$ and the Landau frame constraints:
\bs
\allowdisplaybreaks
\label{eq:aniso_matching}
\beal
  \delta\tilde\ene &= \sum_n \int_p (u\cdot p)^2 \delta\tilde{f}_n = 0 \,,
\\
  \delta\tilde{\mathcal{P}}_L &= \sum_n \int_p (-Z\cdot p)^2 \delta\tilde{f}_n = 0 \,,
\\
  \delta\tilde{\mathcal{P}}_\perp &= \frac{1}{2}\sum_n \int_p (-p \cdot \Xi \cdot p) \delta\tilde{f}_n = 0 \,,
\\
  Q_L &= \sum_n \int_p (u\cdot p)(-Z\cdot p) \delta\tilde{f}_n = 0 \,,
\\
  Q^\mu_\perp &= \sum_n \int_p (u\cdot p) p^{\{\mu\}} \delta\tilde{f}_n = 0 \,,
\end{align}
\es
where $Q_L = {\,-\,} Z {\,\cdot\,} Q$ is the LRF longitudinal heat flow and $Q_\perp^\mu = \Xi^\mu_\nu Q^\nu$ is the transverse heat current. 

\subsection{Modifying the anisotropic distribution}
\label{sec4b}

\begin{figure*}[t]
\includegraphics[width=0.97\linewidth]{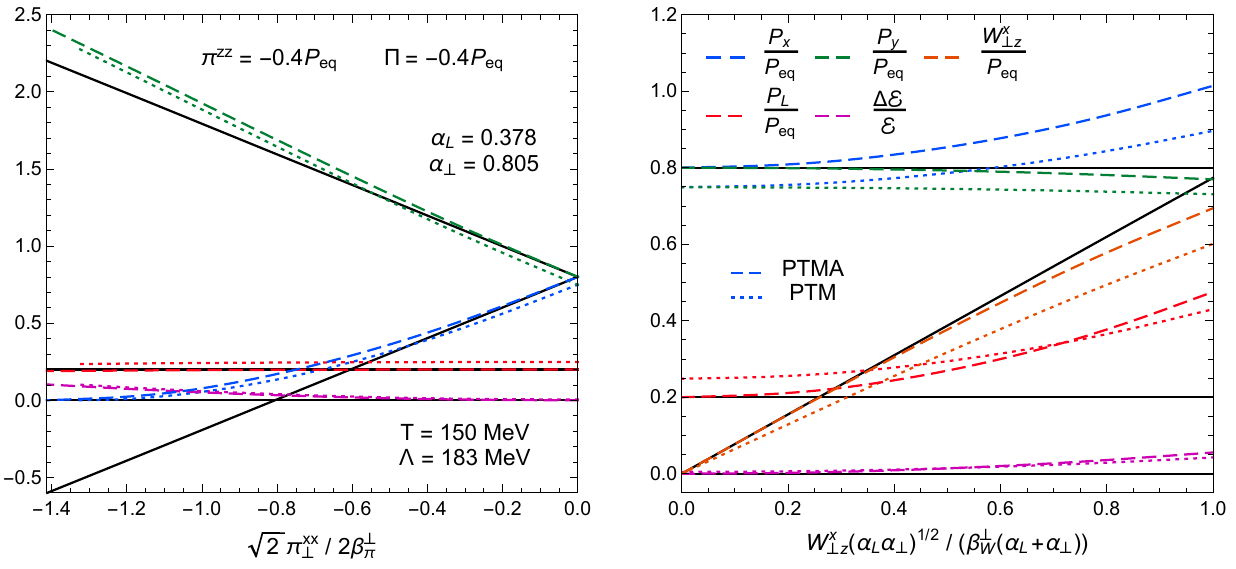}
\caption{
    The reproduction of selected components of the energy-momentum tensor $T^\munu$, using either the PTMA distribution (\ref{eq:famod1}) (dashed color) or the PTM distribution (\ref{eq:feqmod_2}) (dotted color), for a stationary hadron resonance gas at temperature $T = 150$ MeV with zero net baryon density ($\alpha_B = 0$). The system is assumed to have a fixed pressure anisotropy $\pi^{zz} = -\frac{2}{5}\Peq$ and bulk viscous pressure $\Pi = -\frac{2}{5}\Peq$ and is further subjected to either transverse shear stress (left panel) or longitudinal momentum diffusion (right panel). The solid black lines show the components of the hydrodynamic input $T^\munu$. 
\label{famod_output}
}
\end{figure*}

With the anisotropic RTA Chapman--Enskog expansion \eqref{eq:ACE2} at hand we can proceed to modify the leading order anisotropic distribution~\eqref{eq:fa2}. First, we insert a perturbation $\delta\Omega$ in $f_{a,n}$, keeping $\delta\Lambda = 0$:
\be
\label{eq:famod_approx}
  f_{a,n}^{\text{(mod)}} = \frac{g_n}{\exp\biggl[\dfrac{\sqrt{p {\,\cdot\,}(\Omega {+} \delta\Omega) {\,\cdot\,} p {\,+\,} m_n^2}}{\Lambda}\biggr] + \Theta_n}\,.
\ee
Linearizing this modified anisotropic distribution and comparing it to Eq.~\eqref{eq:ACE2} one finds
\be
\label{eq:famod_pert}
  p \cdot \delta \Omega \cdot p = -\frac{p_{\{\mu} p_{\nu\}} \piperp^\munu}{\beta_\pi^\perp \alpha_\perp^2} + \frac{2(-Z \cdot p)p_{\{\mu\}} \Wperp^\mu}{\beta_W^\perp \alpha_\perp \alpha_L}\,.
\ee
Next we rewrite Eq.~\eqref{eq:famod_approx} as\footnote{%
    The generalization of Eq.~\eqref{eq:famod1} for nonzero baryon chemical potential and diffusion is non-trivial and will be left to future work.}
\be
\label{eq:famod1}
  f^{\text{PTMA}}_{a,n} = \frac{\mathcal{Z} g_n}{\exp\biggl[\dfrac{\sqrt{\bm{p}''^2 {+} m_n^2}}{\Lambda}\biggr] + \Theta_n}\,,
\ee
where the normalization $\mathcal{Z}$ will be discussed further below and
\be
  p_i = B_{ij} p^{\prime\prime}_j = C_{im} A_{mj} p^{\prime\prime}_j \,.
\ee
By comparing with Eq.~\eqref{eq:Aij_mod} one sees that the residual shear transformation $C_{im}$ further deforms the anisotropic momentum space. The additional deformations, which are linearly proportional to $\piperp^\munu$ and $\Wperp^\mu$, are assumed to be smaller than the anisotropy parameters $\alpha_L$ and $\alpha_\perp$. The matrix $C_{im}$ is constructed such that the total deformation matrix $B_{ij} = C_{im} A_{mj}$ is symmetric and reproduces the perturbation $\delta\Omega$ in Eq.~\eqref{eq:famod_approx} for small residual shear stresses. After some algebra one finds
\be
\label{eq:Cmatrix}
  C_{im} = \delta_{im} + \frac{\pi_{\perp,im}}{2\beta_\pi^\perp} + \frac{\alpha_\perp W_{\perp z, i} Z_m + \alpha_L W_{\perp z, m} Z_i}{\beta_W^\perp(\alpha_\perp {+} \alpha_L)}\,.
\ee
Here $\pi_{\perp,im} = X_i {\,\cdot\,} \pi_\perp {\cdot\,} X_m$ and $W_{\perp z,i} = - X_i {\,\cdot\,} \Wperp$ are the LRF residual shear stress components. Although $C_{im}$ in Eq.~\eqref{eq:Cmatrix} is not symmetric, $B_{ij}$ is:
\be
\label{eq:Bij}
  B_{ij} = A_{ij} + \frac{\alpha_\perp \pi_{\perp,ij}}{2\beta_\pi^\perp} + \frac{\alpha_\perp \alpha_L (W_{\perp z, i} Z_j + W_{\perp z, j} Z_i)}{\beta_W^\perp(\alpha_\perp {+} \alpha_L)}.
\ee
Finally we renormalize the particle density to the one given by the leading order anisotropic distribution:
\be
  n_{a,n} = {\det} A \cdot n_{\eq,n}(\Lambda,0)\,
\ee
where $n_{a,n}$ is the anisotropic particle density and ${\det}A = \alpha_\perp^2\alpha_L$ (there is no contribution from the anisotropic RTA Chapman--Enskog correction $\delta\tilde{f}_n$). The modified particle density from Eq.~\eqref{eq:famod1} is
\be
 n_{a,n}^{\text{PTMA}} = \mathcal{Z} \, {\det} C \cdot n_{a,n} \,,
\ee
which leads us to the normalization factor 
\be
\label{eq:Z_mod}
  \mathcal{Z} = \frac{1}{{\det} C}\,.
\ee

In Figure~\ref{famod_output}, we test the reproduction of the energy-momentum tensor by the ``PTMA distribution" \eqref{eq:famod1}, similar to the test done in Fig.~\ref{hydro_output}. We consider a hadron resonance gas at temperature $T = 150$ MeV and baryon chemical potential $\alpha_B = 0$ with an input $T^\munu$ featuring the two fixed dissipative flows $\pi^{zz} = \Pi = -\frac{2}{5}\Peq$ (or, equivalently, $\PL = \frac{1}{5}\Peq$ and $\Pperp = \frac{4}{5}\Peq$). These viscous pressures are captured by the leading order distribution $f_{a,n}$ by numerically adjusting (``Landau matching'') the anisotropy parameters $\Lambda$, $\alpha_L$ and $\alpha_\perp$ accordingly. 

In the left panel of Fig.~\ref{famod_output} we explore the reliability of the modified anisotropic distribution in reproducing a nonzero transverse shear stress $\piperp^{xx} = -\piperp^{yy}$. Moving from the right edge of the plot leftward we decrease $\piperp^{xx}$, causing the pressures $\mathcal{P}_x = \Pperp + \piperp^{xx}$ and $\mathcal{P}_y = \Pperp + \piperp^{yy}$ to decrease and increase, respectively. According to the anisotropic matching conditions \eqref{eq:aniso_matching}, the energy density and longitudinal pressure should remain constant as we do so. The hydrodynamic input (solid black lines) and kinetic output $T^\munu$ components (colored dashed lines) from this test are very similar to those shown in the left panels of Fig.~\ref{hydro_output}, with the kinetic output for $\mathcal{P}_x$ (blue dashed) approaching zero for large negative $\piperp^{xx}$ while the output for  $\mathcal{P}_y$ (green dashed) overestimates the hydrodynamic target value but to a lesser degree. The kinetic outputs for $\ene$ and $\PL$ are in very good agreement with their hydrodynamic target values. The modified anisotropic distribution breaks down for $\piperp^{xx} < -2\beta_\pi^\perp$ (or $\det C < 0$) but for moderately large values $|\piperp^{xx}| \leq \frac{1}{6}(\PL {+} 2\Pperp)$ the discrepancies between the input and output values are $|\Delta \mathcal{P}_x|/\mathcal{P}_x \leq 4.0 \%$, $|\Delta \mathcal{P}_y|/\mathcal{P}_y \leq 1.5 \%$, $|\Delta \PL|/\PL \leq 0.17 \%$ and $|\Delta \ene|/\ene \leq 0.47 \%$. We also checked that for this hydrodynamic input the off-diagonal components of the kinetic output $T^\munu$ are zero. 

For the second test (right panel of Fig.~\ref{famod_output}) we start from the same system but add a longitudinal momentum diffusion component $W_{\perp z}^x$ of increasing magnitude. The energy density and pressure components $\mathcal{P}_x$, $\mathcal{P}_y$ and $\PL$ should not change as we do so. Overall, we find that the kinetic outputs for $\ene$, $\mathcal{P}_y$ and $W_{\perp z}^x$ are in good agreement with this expectation. However, the kinetic outputs for the pressure components $\mathcal{P}_x$ and $\PL$ are seen to exhibit stronger sensitivity to the errors caused by mismatched nonlinear terms in $W_{\perp z}^x$. Still, for moderate values of $|W_{\perp z}^x| \leq \frac{1}{6}(\PL {+} 2\Pperp)$, the output errors stay below $|\Delta W_{\perp,z}^x / W_{\perp,z}^x| \leq 1.6 \%$, $|\Delta \mathcal{P}_x|/\mathcal{P}_x \leq 4.0 \%$, $|\Delta \mathcal{P}_y|/\mathcal{P}_y \leq 0.55 \%$, $|\Delta \PL|/\PL \leq 21 \%$ and $|\Delta \ene|/\ene \leq 0.84 \%$

Finally, we repeat the previous tests with the PTM modified equilibrium distribution~\eqref{eq:feqmod_2} and plot 
the corresponding kinetic output $T^\munu$ as dotted lines in Fig.~\ref{famod_output} for comparison. One observes that the PTM distribution follows essentially the same trends as the PTMA distribution, but it does not fully capture the pressure anisotropy and bulk viscous pressure at zero residual shear stress, similar to what we saw in Fig.~\ref{hydro_output}. On the other hand, by imposing the generalized Landau matching conditions (\ref{eq:aniso_matching}b,c) the leading order anisotropic distribution precisely reproduces the longitudinal and transverse pressures in that limit.

\begin{figure*}[t]
\centering
\includegraphics[width=\linewidth]{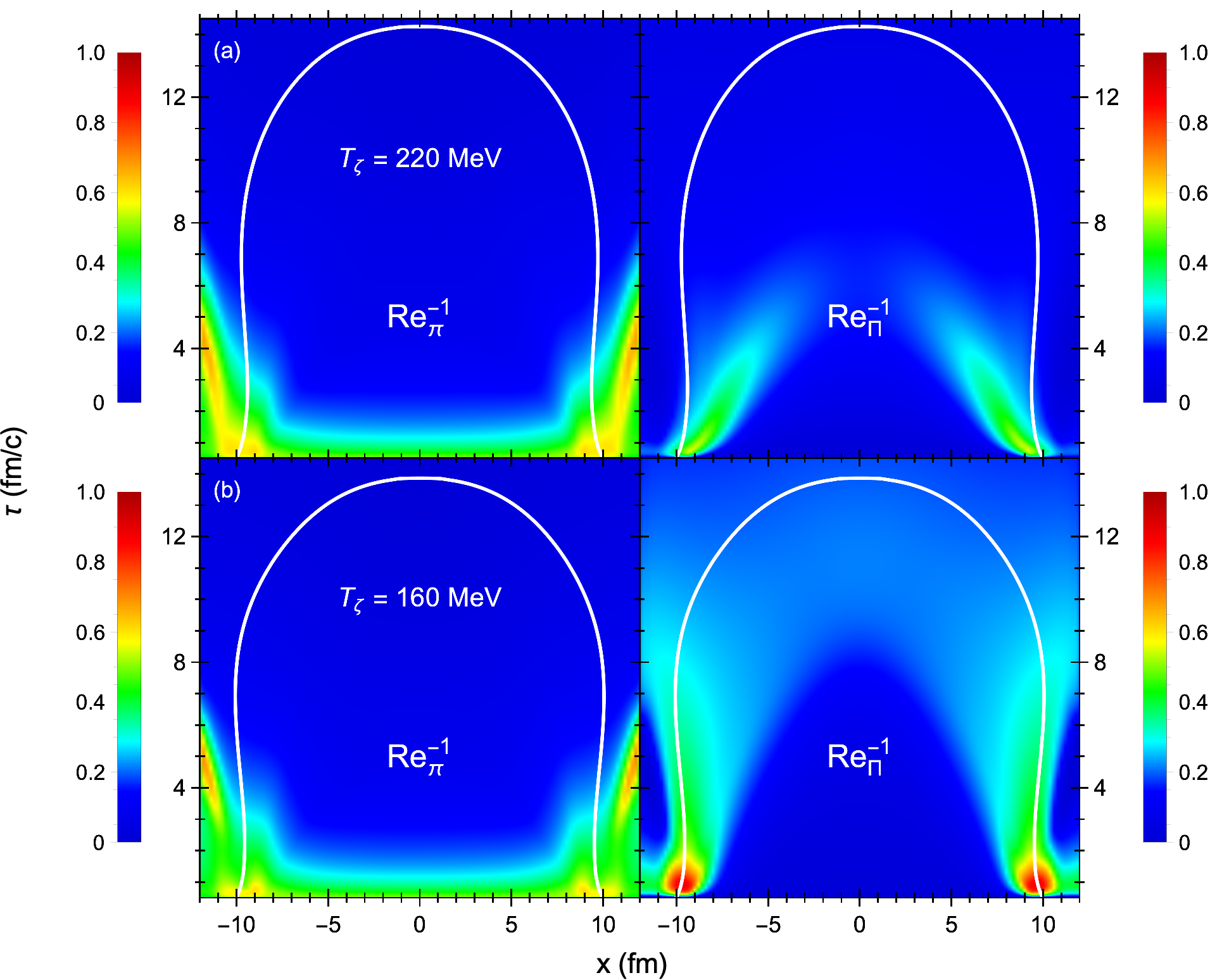}
\caption{
    The $\tau{-}x$ slice $(y = \eta_s = 0)$ of the shear inverse Reynolds number $\text{Re}^{-1}_\pi = \sqrt{\pi^\munu \pi_\munu} / (\Peq\sqrt{3})$ (left panels) and bulk inverse Reynolds number $\text{Re}^{-1}_\Pi = |\Pi| / \Peq$ (right panels) showing the strength of the $\delta f_n$ corrections on a particlization hypersurface of constant temperature $T_\text{sw} = 150$ MeV (white contour) from a (2+1)--d central Pb+Pb collision with smooth \trento{} initial conditions.  We vary the peak temperature of the specific bulk viscosity $\zeta / \mathcal{S}$ to either $T_\zeta = 220$ MeV (top row) or $T_\zeta = 160$ MeV (bottom row).
\label{feqmod_hydro_slice}
}
\end{figure*}

\section{Continuous particle spectra}
\label{sec5}

In this section we compute the continuous momentum spectra of identified hadrons $(\pi^+, K^+, p)$ using the Cooper--Frye formula \eqref{eq:CooperFrye}. To generate the hypersurfaces, we run the {\sc VAH} code \cite{McNelis:2021zji} to evolve central and non-central Pb+Pb collisions using standard viscous hydrodynamics with smooth initial conditions. For simplicity, we only consider the central slice ($\eta_s=0$) in the transverse plane and assume longitudinal boost-invariance to extend the solution in the spacetime rapidity direction. For the different hadron phase-space distribution models discussed in Secs.~\ref{sec2} -- \ref{sec4}, we will compare the azimuthally averaged transverse momentum spectra 
\be
\label{eq:pT_spectra}
  \frac{dN_n}{2\pi p_T dp_T dy_p} = \int_0^{2\pi}\, \frac{d\phi_p}{2\pi}\frac{dN_n}{p_T dp_T d\phi_p dy_p}
\ee
and the $p_T$--differential elliptic flow coefficient
\be
\label{eq:v2}
  v^{(2)}_2(p_T) = \frac{\int_0^{2\pi} d\phi_p \cos(2\phi_p)\,\dfrac{dN_n}{p_T dp_T d\phi_p dy_p}}{\int_0^{2\pi} d\phi_p\, \dfrac{dN_n}{p_T dp_T d\phi_p dy_p}}
\ee
at mid-rapidity ($y_p = 0$). Specifically, we compare results obtained with the 14--moment approximation \eqref{eq:14_moment_irreducible}, the RTA Chapman--Enskog expansion \eqref{eq:Chapman_Enskog}, the PTM and PTB modified equilibrium distributions \eqref{eq:feqmod_2} and \eqref{eq:Jonah}, and the PTMA modified anisotropic distribution \eqref{eq:famod1}. 

\vspace*{-2mm}
\subsection{Setup}
\label{sec5a}
\vspace*{-3mm}

We evolve an azimuthally symmetric, event-averaged \trento{} transverse energy density profile with (2+1)--dimensional second-order viscous hydrodynamics~\cite{Moreland:2014oya, McNelis:2021zji}. We start the simulation at the longitudinal proper time $\tau_0 = 0.5$\,fm/$c$ with an initial central temperature of $T_{0,\text{center}} = 400$ MeV. The spatial components of the fluid velocity $u^\mu$ and bulk viscous pressure $\Pi$ are initialized to zero. The shear stress tensor is initialized as
\be
    \pi^\munu = \frac{1}{3}\big(\PL {-} \Pperp\big) \left(\Delta^\munu + 3Z^\mu Z^\nu\right)\,,
\ee
where the initial pressure anisotropy is set to $\PL - \Pperp = \frac{12}{11}\Peq$ and the initial longitudinal basis vector is $Z^\mu = (0,0,0,\tau_0^{-1})$. For the equilibrium pressure $\mathcal{P}_\eq(\ene)$, we use the lattice QCD equation of state from the HotQCD collaboration~\cite{Bazavov:2014pvz}. The baryon chemical potential $\alpha_B$ and baryon diffusion current $V_B^\mu$ are fixed to zero for the entire simulation.\footnote{%
    The current version of the relativistic hydrodynamic code {\sc VAH} \cite{McNelis:2021zji} does not propagate the net baryon density and baryon diffusion current.} 

The shear and bulk viscosities are modeled using the temperature dependent parametrizations from the JETSCAPE collaboration \cite{Everett:2020xug,Everett:2020yty}:
\bs
\allowdisplaybreaks
\begin{align}
    (\eta/\mathcal{S})(T) =&\, (\eta/\mathcal{S})_\text{kink} \,+\, a_\text{low}\,(T {-} T_\eta) \Theta(T_\eta {-} T) \\ \nonumber
    & \,+\, a_\text{high}(T {-} T_\eta)\,\Theta(T {-} T_\eta)\,, \\
    (\zeta/\mathcal{S})(T) =&\, \frac{(\zeta/\mathcal{S})_\text{max} \,\Lambda_\zeta(T)^2}{\Lambda_\zeta(T)^2 + (T {-} T_\zeta)^2}\,,
\end{align}
\es
where $\Theta(x)$ is the Heaviside step function and
\be
    \Lambda_\zeta(T) = w_\zeta \left(1 + \lambda_\zeta \,\mathrm{sgn}(T{-}T_\zeta)\right)\,.
\ee
In this work we fix the viscosity parameters to $(\eta/\mathcal{S})_\text{kink} = 0.093$, $T_\eta = 0.22$ GeV, $a_\text{low} = -0.77$ GeV$^{-1}$, $a_\text{high} = 0.21$ GeV$^{-1}$, $(\zeta/\mathcal{S})_\text{max} = 0.1$, $w_\zeta = 0.05$ GeV and $\lambda_\zeta = 0$. However, for exploration purposes we vary the temperature $T_\zeta$ at which $\zeta / \mathcal{S}$ peaks, setting it to either $220$ MeV or $160$ MeV. 

From the hydrodynamic simulation we generate an isothermal particlization hypersurface of temperature $T_\text{sw} = 150$ MeV, using the freeze-out finder code {\sc CORNELIUS} \cite{Huovinen:2012is}. For our central Pb+Pb collision, the longitudinally boost-invariant hypersurface at spacetime rapidity $\eta_s = 0$ contains about $6.8 {\,\times\,} 10^4$ freeze-out cells. Figure~\ref{feqmod_hydro_slice} shows a $\tau{-}x$ slice of the particlization surface at $y = \eta_s = 0$, as well as the shear and bulk inverse Reynolds numbers to gauge the strength of the shear and bulk $\delta f_n$ corrections.\footnote{%
    We note the setup used in this Section is different from the one in an earlier study to test the {\sc iS3D} particle sampler (see Sec.~3 in Ref.~\cite{McNelis:2019auj}). There we found that the viscous corrections were so large that the PTMA distribution was unusable for a significant part of the hypersurface. The hypersurface shown in Fig.~\ref{feqmod_hydro_slice} was computed with parameter settings that allow for a meaningful comparison between the PTM and PTMA distributions.
    \label{fn14}} 
Finally, we evaluate the Cooper--Frye formula with the code {\sc iS3D}~\cite{McNelis:2019auj}, which was developed from the particlization module {\sc{iSS}} \cite{Shen:2014vra}. The code gives the user the option to use one of five $\delta f_n$ corrections described in Secs.~\ref{sec2} -- \ref{sec4}. To compute the longitudinally boost-invariant particle spectra, the hypersurface volume needs to be extended to the $\eta_s$ dimension and centered around the momentum rapidity $y_p$. We perform the numerical integration of the Cooper--Frye formula along the $\eta_s$--direction using Gauss--Legendre integration on a 48-point grid $(y_p {\,-\,} \eta_s)_j$ with integration weights $\omega_j$ given by
\bs
\allowdisplaybreaks
\label{eq:integration}
\beal
  (y_p {\,-\,} \eta_s)_j &= \sinh^{-1}\left(\frac{x_j}{1{\,-\,}x_j^2}\right) \,,
\\
  \omega_j &= w_j \frac{1{\,+\,}x_j^2}{|1{\,-\,}x_j^2|\sqrt{1{\,-\,}x_j^2{\,+\,}x_j^4}}\,,
\end{align}
\es
where $x_j$ and $w_j$ are the Gauss--Legendre roots and weights, respectively.

\subsection{Breakdown of the modified distributions and technical issues}
\label{sec5b}

As stated in the previous sections, the modified equilibrium distributions \eqref{eq:feqmod_2} and \eqref{eq:Jonah} break down in freeze-out cells with large viscous corrections. For the first scenario where the specific bulk viscosity's peak temperature is $T_\zeta = 220$ MeV (top row in Fig.~\ref{feqmod_hydro_slice}), both the shear and bulk viscous corrections are small enough to use the modified equilibrium distribution for all freeze-out cells. For $T_\zeta = 160$ MeV (bottom row in Fig.~\ref{feqmod_hydro_slice}) the bulk viscosity peaks much closer to the particlization hypersurface. Not only are the bulk viscous corrections much larger in this case, but the shear corrections are also enhanced due to shear-bulk coupling effects in the hydrodynamic simulation \cite{Denicol:2014mca}. Together they cause the modified equilibrium distribution to break down for about 4400 freeze-out cells at $r{\,\sim\,}10$\,fm and $0.5$\,fm/$c{\,<\,}\tau{\,<\,}1.3$\,fm/$c$. There are several options to handle the particle spectra contributions from these freeze-out cells: (i) ignore such cells entirely, (ii) use for them the local-equilibrium distribution $f_n = f_{\eq,n}$, which is positive definite but neglects the viscous components of $T^\munu$, or (iii) linearize in such cells the modified equilibrium distribution $f_{\eq,n}^{(\text{mod})} \approx f_{\eq,n} + \delta f_n$, which may turn negative at high momenta but captures all components of $T^\munu$. Here we choose the third option, i.e. for the PTM distribution we switch to the Chapman--Enskog expansion~\eqref{eq:Chapman_Enskog} when ${\det} A < 10^{-5}$ or when the normalization factor $\mathcal{Z}_n$ of any hadron (usually the lightest pion $\pi^0$) turns negative. For the PTB distribution, we use $\delta f_n$ from Eq.~\eqref{eq:linear_Jonah} when ${\det} A < 10^{-5}$ or $-\Peq < \Pi < \ene/3 - \Peq$. In this study, the impact of these manipulations on the total particle spectra is negligible since the number of ``bad'' freeze-out cells is small. However, if a significant fraction of the hypersurface requires one to use something other than the modified equilibrium distribution, then Eqs.~\eqref{eq:feqmod_2} and \eqref{eq:Jonah} should probably not be used for particlization.

\begin{figure*}[t]
\centering
\includegraphics[width=\linewidth]{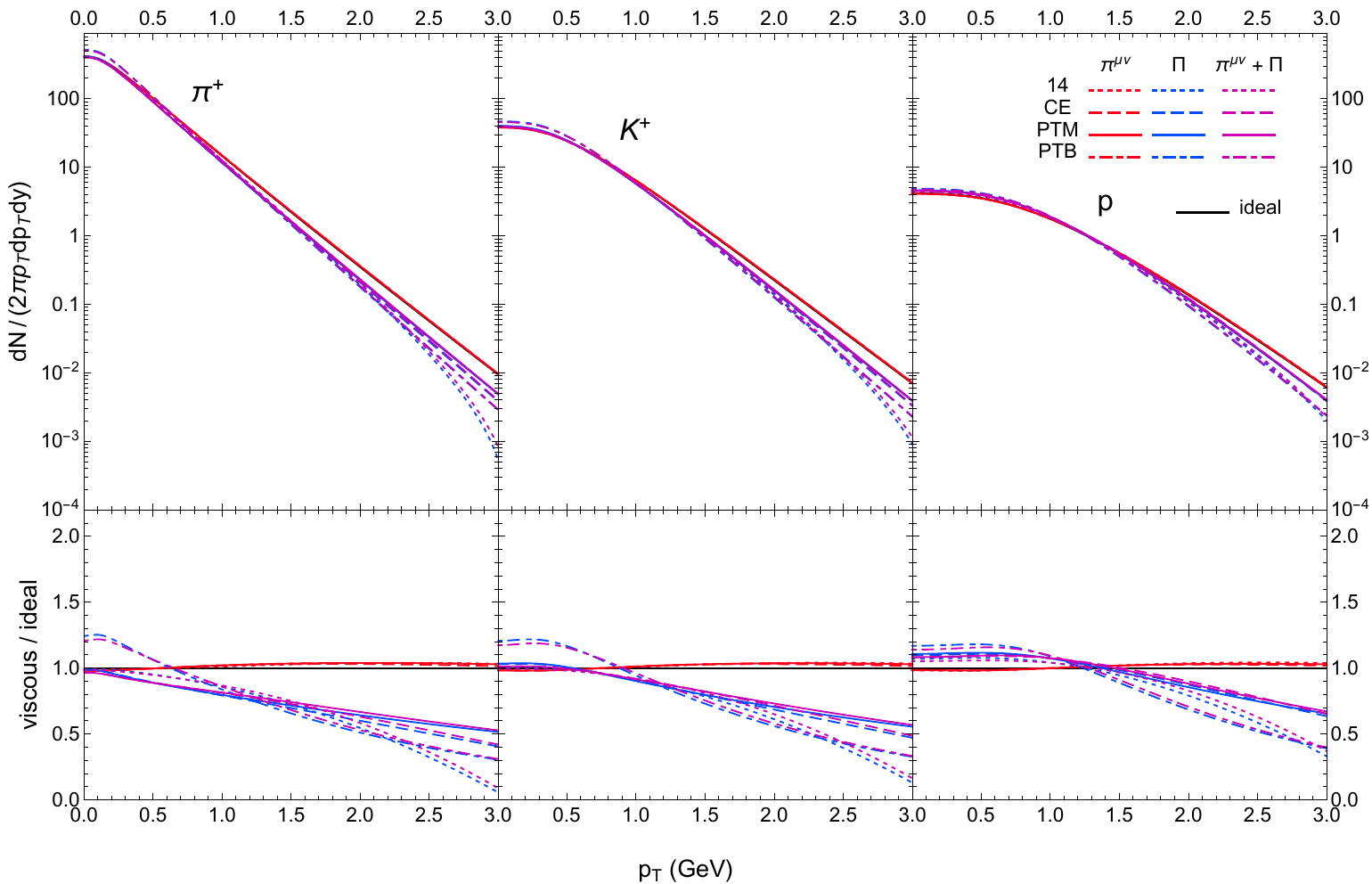}
\caption{
    The azimuthally-averaged transverse momentum spectra of $(\pi^+, K^+, p)$ (top panels) computed with the Cooper--Frye formula for the (2+1)--d central Pb+Pb collision described in Sec.~\ref{sec5a}. The peak temperature of $\zeta/\mathcal{S}$ is set to $T_\zeta = 220$ MeV. We compare the shear (red), bulk (blue) and combined shear + bulk (purple) $\delta f_n$ corrections of the 14--moment approximation (dotted color), RTA Chapman--Enskog expansion (dashed color), PTM distribution (solid color) and PTB distribution (dot-dashed color) relative to the \textit{ideal} spectra with $\delta f_n = 0$ (solid black). The bottom panels show the ratio of the particle spectra with $\delta f_n$ corrections to the \textit{ideal} spectra. 
\label{dNdpT_small_bulk}
}
\end{figure*}

Even if the modified equilibrium distribution can be used, a very small ${\det}A$ value can cause its width along the rapidity direction $y_p {\,-\,} \eta_s$ to be extremely narrow. This leads to numerical errors in the particle yields if the spacetime rapidity grid $(y_p {\,-\,} \eta_s)_j$ is not fine enough. To resolve this technical issue, we rescale the spatial grid $(y_p {\,-\,} \eta_s)_j$ by the distribution's rapidity width $\delta y_p$. One can estimate $\delta y_p$ from a diagonal deformation matrix $A_{ij} {\,=\,} \diag(1{\,+\,}\bar\Pi{\,-\,}\frac{1}{2}\bar{\pi}_{zz},1{\,+\,}\bar\Pi{\,-\,}\frac{1}{2}\bar{\pi}_{zz},1{\,+\,}\bar\Pi{\,+\,}\bar{\pi}_{zz})$:
\be
    \delta y_p \sim 1+\bar\Pi+\bar{\pi}_{zz}\,,
\ee
where $\bar\Pi = \Pi/(3\beta_\Pi)$ (or $\lambda_\Pi$) and $\bar{\pi}_{zz} = \pi_{zz}/(2\beta_\pi)$. Relative to the transverse momentum space, the rapidity distribution becomes very narrow when $\bar{\pi}_{zz} \approx - (1+\bar\Pi)$. In this limit, the rapidity width is proportional to
\be
    \delta y_p \propto \frac{{\det}A}{({\det}A_\Pi)^{2/3}}\,,
\ee
where ${\det}A_\Pi = (1{\,+\,}\bar\Pi)^3$. Therefore, for each longitudinally boost-invariant freeze-out cell $d^2\sigma_{\mu,i}$ we rescale the spacetime rapidity grid points and weights~\eqref{eq:integration} as
\bs
\allowdisplaybreaks
\label{eq:integration_rescale}
\beal
    (y_p {\,-\,} \eta_s)_{i,j} &= \frac{{\det}A_i}{({\det}A_{\Pi,i})^{2/3}} \times (y_p {\,-\,} \eta_s)_j \,,
\\
    \omega_{i,j} &= \frac{{\det}A_i}{({\det}A_{\Pi,i})^{2/3}} \times \omega_j\,.
\end{align}
\es
This rescaling trick is found to work well even for small values of ${\det}A \sim 10^{-5}$. For (3+1)--d hypersurfaces, however, this method cannot be used because the freeze-out finder fixes the freeze-out cells' spacetime rapidity. Instead, we switch to a linearized $\delta f_n$ correction if the rapidity width is too small (e.g. ${\det}A / ({\det}A_\Pi)^{2/3} < 0.01$).

The modified anisotropic distribution~\eqref{eq:famod1} can also break down, usually because the longitudinal pressure $\PL$ turns negative during the viscous hydrodynamic simulation. When this happens one cannot construct a solution for the momentum deformation parameter $\alpha_L$ in the leading order anisotropic distribution \eqref{eq:fa}. For the hypersurface with large bulk viscous corrections we find $\PL{\,<\,}0$ in about 7800 freeze-out cells during the period $0.5$\,fm/$c < \tau < 2.3$\,fm/$c$ (see footnote \ref{fn14}). For these freeze-out cells, we here simply replace Eq.~\eqref{eq:famod1} by the local-equilibrium distribution \eqref{eq:feq}. In practice it would be more appropriate to use the PTMA distribution on hypersurfaces constructed from {\it anisotropic} hydrodynamic simulations in which the occurrence of negative longitudinal pressures is largely avoided \cite{McNelis:2021zji}.

\begin{figure*}[t]
\centering
\includegraphics[width=\linewidth]{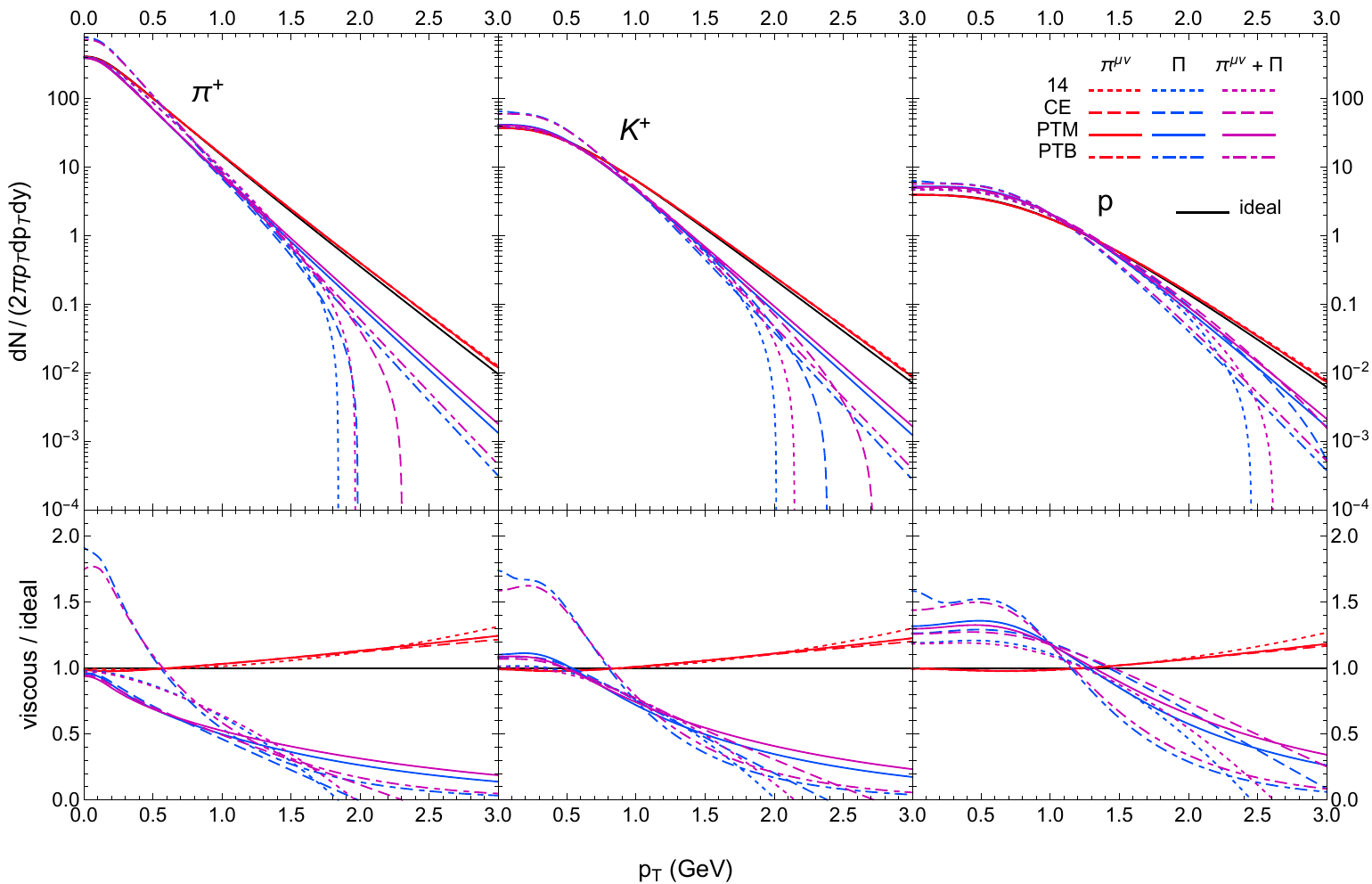}
\caption{
    Same as Fig.~\ref{dNdpT_small_bulk} but with a $\zeta / \mathcal{S}$ peak temperature of $T_\zeta = 160$ MeV. 
\label{dNdpT_large_bulk}
}
\end{figure*}

\subsection{Central collisions}
\label{sec5c}

Figure~\ref{dNdpT_small_bulk} shows the continuous transverse momentum spectra \eqref{eq:pT_spectra} of ($\pi^+$, $K^+$, $p$), without resonance decay contributions or hadronic rescattering, for our central Pb+Pb collision, using a specific bulk viscosity peak temperature of $T_\zeta = 220$ MeV. We study the shear and bulk viscous corrections to the \textit{ideal} spectra (i.e. $\delta f_n = 0$) computed with the four models for $\delta f_n$ discussed in Sec.~\ref{sec2}.

As seen in the figure, the shear stress (although small) slightly flattens the $p_T$ spectra (red curves), without affecting the total yields, since it slows the longitudinal expansion and pushes the particles outward in the transverse direction. The bulk viscous pressure has the opposite effect by counteracting the scalar expansion rate and reducing the average pressure, thereby softening the $p_T$ spectra (blue curves). One observes the ``shoulder'' at low values of $p_T$ that is typical for thermal flow spectra \cite{Schnedermann:1993ws}; there the bulk corrections are more pronounced in protons than in pions and kaons. Furthermore, the bulk viscous correction decreases the total pion and kaon yields while increasing the proton yield. The PTB distribution is the only exception to these trends: for non-zero bulk viscous pressure it both raises the shoulder and increases the yields of all particles, which sets it apart from the other $\delta f_n$ corrections. Finally, the purple curves show the combined effects of the shear and bulk viscous corrections to the spectra. Overall, the slope of the total $p_T$ spectra is steeper than the \textit{ideal} one since here the bulk viscous pressure is larger than the shear stress for most of the hypersurface.

We also compare the $\delta f_n$ corrections of the 14--moment approximation, RTA Chapman--Enskog expansion, PTM distribution and PTB distribution. We see that the momentum dependence of the viscous correction varies considerably between models: the ratio of the total particle spectra to the $\textit{ideal}$ one at high $p_T$ is approximately quadratic for the 14--moment approximation and linear for the RTA Chapman--Enskog expansion, as illustrated by the bottom panels in Fig.~\ref{dNdpT_small_bulk}. The PTM spectra are almost identical to the ones computed with the Chapman--Enskog expansion because the shear stress and bulk viscous pressures are small; differences between the two approaches emerge only at large values of $p_T$. Overall, there are no significant differences among the first three $\delta f_n$ models across most of the $p_T$ spectrum in Fig.~\ref{dNdpT_small_bulk}. The PTB distribution, on the other hand, has a moderate excess of low $p_T$ pions and kaons relative to the other $\delta f_n$ corrections.

\begin{figure*}[t]
\centering
\includegraphics[width=\linewidth]{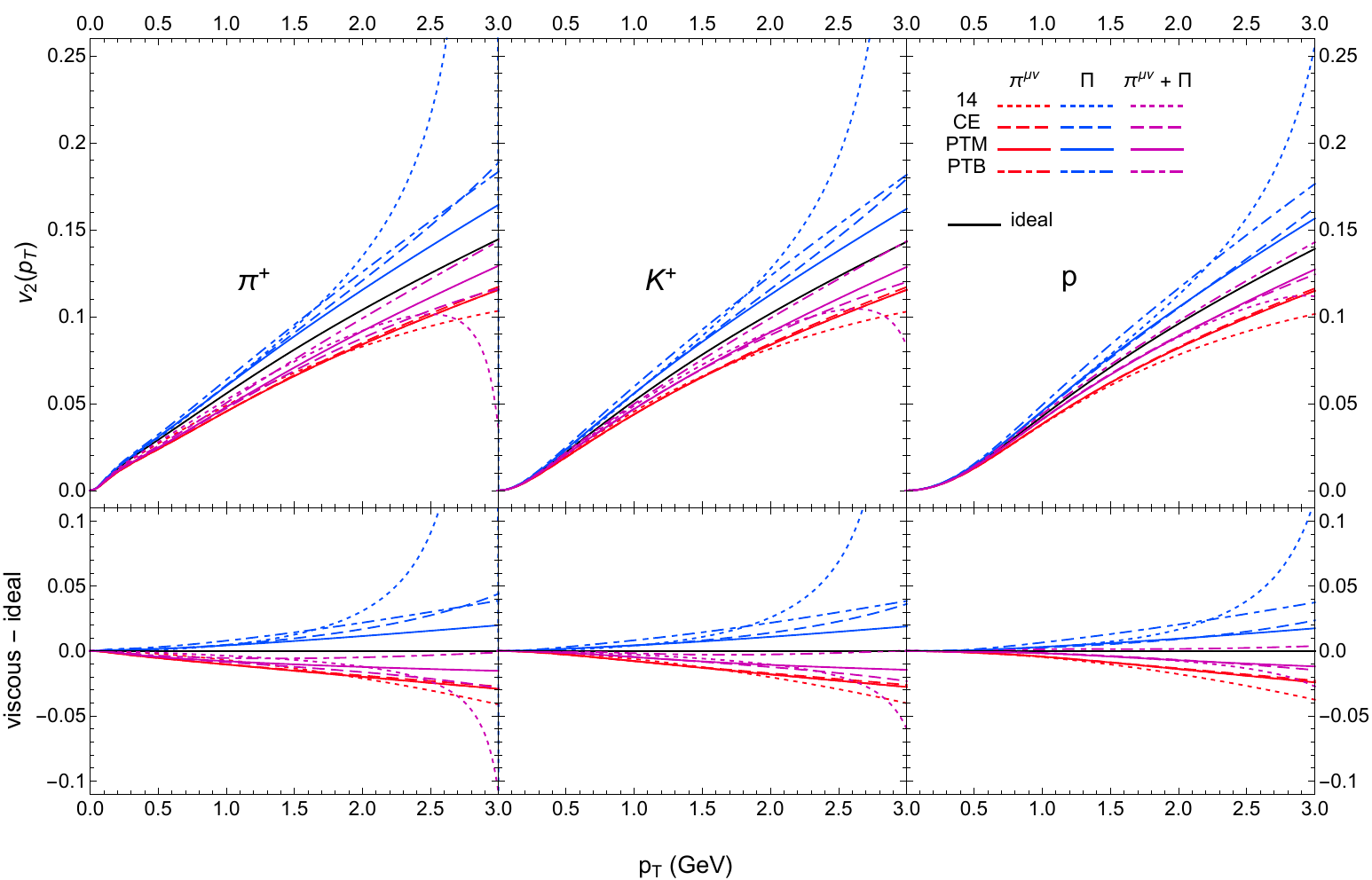}
\caption{(Color online)
    The $p_T$--differential elliptic flow coefficient of $(\pi^+, K^+, p)$ for the (2+1)--d non-central Pb+Pb collision (top panels). The peak temperature of $\zeta / \mathcal{S}$ is set to $T_\zeta = 220$ MeV. Similar to Fig.~\ref{dNdpT_small_bulk}, we plot the shear and bulk viscous corrections of each $\delta f_n$ model to the \textit{ideal} $v_2(p_T)$ (solid black). The bottom panels show the differences between $v_2(p_T)$ with $\delta f_n$ corrections and the \textit{ideal} $v_2(p_T)$.
\label{v2_small_bulk}
}
\end{figure*}

In Figure~\ref{dNdpT_large_bulk}  we show how the central Pb+Pb collision spectra from Fig.~\ref{dNdpT_small_bulk} change when we move the peak of the specific bulk viscosity from $T_\zeta = 220$ MeV to 160 MeV. This obviously increases the magnitude of the bulk viscous pressure on the particlization hypersurface at $T_\mathrm{sw}=150$ MeV. The shear-bulk coupling effect in the hydrodynamic simulation then also increases the strength of the shear viscous corrections on the hypersurface, resulting in much flatter $p_T$ slopes than in the previous case. Surprisingly, the shear viscous corrections from the four $\delta f_n$ models are still very close to each other (for PTM and PTB this is expected because these distributions use the same shear stress modification.) Instead, the largest differences are found in their bulk viscous corrections. Because the bulk viscous pressure on the hypersurface is quite large (lower right panel in Fig.~\ref{feqmod_hydro_slice}), the linearized bulk viscous correction in the 14--moment approximation causes the pion spectra to turn negative already at intermediate momentum $p_T \sim 1.8$ GeV while the kaon and proton spectra flip sign for $p_T > 2$ GeV and 2.4 GeV, respectively. The bulk viscous correction in the RTA Chapman--Enskog expansion is less severe than in the 14--moment approximation, but the pion and kaon spectra still turn negative for $p_T \gtrsim 2$ GeV (the proton spectra remain positive up to $p_T = 3$ GeV). In contrast, the PTM spectra are positive definite by construction even for moderately large bulk viscous pressures, maintaining an exponential tail at high values of $p_T$.\footnote{%
    Without regulation, the linearized $\delta f_n$ corrections from the ``bad" freeze-out cells (see Sec.~\ref{sec5b}) eventually turn the PTM and PTB spectra negative but this occurs outside the experimental range of interest for soft hadron emission $p_T < 3$ GeV.} 
One also observes a slight excess of protons at low $p_T$ (this effect is much less visible for kaons). The PTB spectra also remain positive but have significantly steeper slopes than the PTM spectra; the shoulder enhancements at low $p_T$ are also much larger, especially for pions.

After combining the shear and bulk viscous corrections, the $p_T$ spectra become positive except for the one computed with the 14--moment approximation, which has a strong quadratic momentum dependence from its bulk viscous correction. 

\begin{figure*}[t]
\centering
\includegraphics[width=\linewidth]{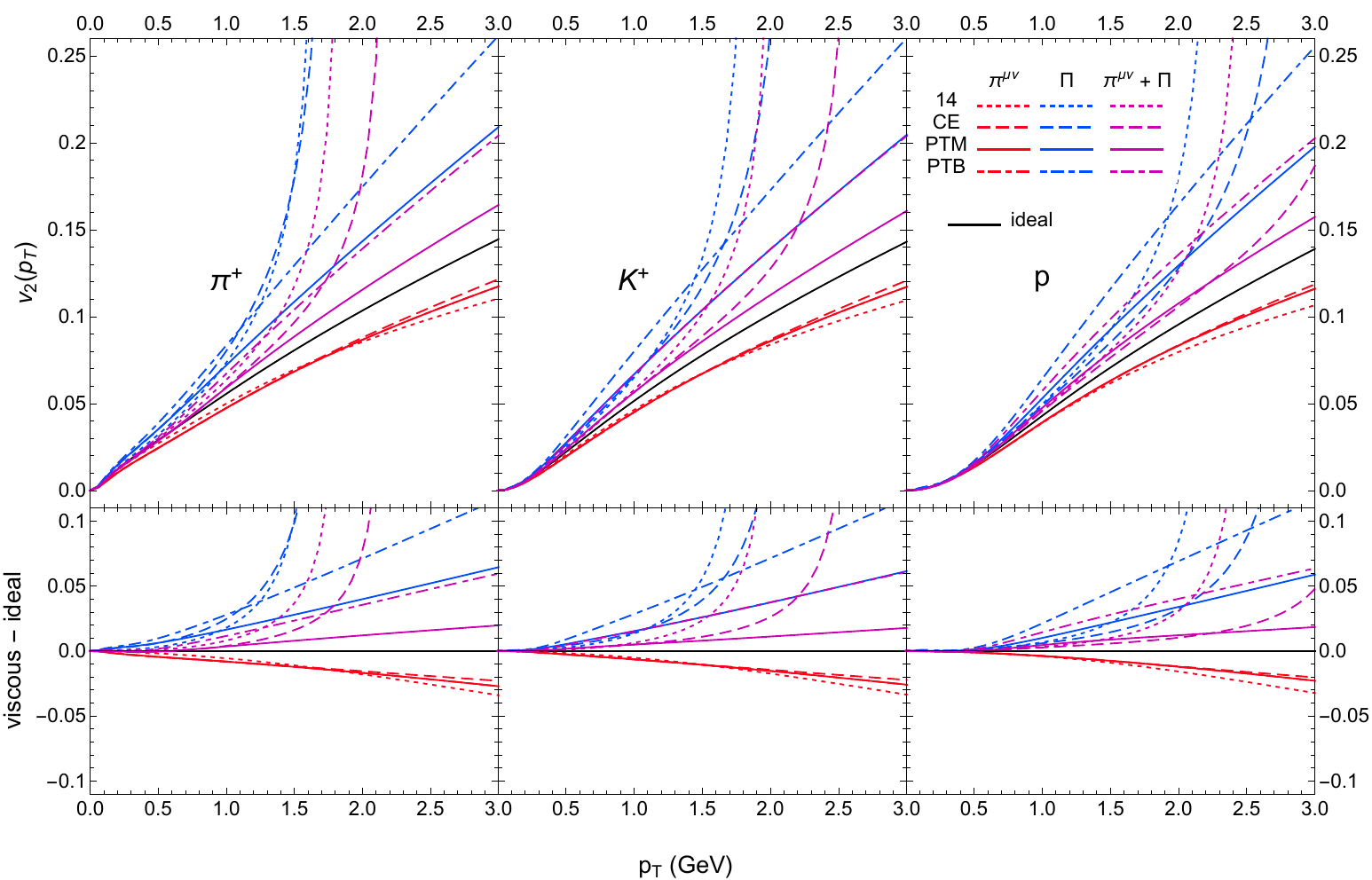}
\caption{
    Same as Fig.~\ref{v2_small_bulk} but with a $\zeta / \mathcal{S}$ peak temperature of $T_\zeta = 160$ MeV. 
\label{v2_large_bulk}
}
\end{figure*}

\subsection{Non-central collisions}
\label{sec5d}

Next we repeat the same hydrodynamic simulations for a nonzero impact parameter $b = 5$\,fm, to study the effects of the different viscous corrections $\delta f_n$ on the $p_T$--differential elliptic flow coefficient, shown in Fig.~\ref{v2_small_bulk} for $\pi^+$, $K^+$, and $p$. Here we start with $T_\zeta = 220$ MeV, resulting in relatively weak viscous stresses on the particlization hypersurface.
The shear viscous corrections are seen to decrease the differential elliptic flow $v_2(p_T)$, counteracting the effects of anisotropic transverse flow. The bulk viscous pressure, on the other hand, tends to increase $v_2(p_T)$ by suppressing the radial flow and making the $p_T$ spectra \eqref{eq:pT_spectra} steeper. Individually, the small shear and bulk viscous corrections to the \textit{ideal} $v_2(p_T)$ (defined by setting $\delta f_n{\,=\,}0$) are roughly linear in $p_T$, except for the bulk $\delta f_n$ correction of the 14--moment approximation, which causes $v_2(p_T)$ to diverge at high $p_T$ when the spectrum (which enters the denominator of $v_n(p_T)$) passes through zero. Overall, there is a net suppression on the differential elliptic flow since it is more sensitive to the shear viscous corrections at the space-like edges of the hypersurface, whose fluid cells have undergone the strongest transverse acceleration. Interestingly, the only exception is the PTB elliptic flow, whose shear and bulk viscous corrections nearly cancel each other. Similar to the spectra discussed in the previous subsection, the four $\delta f_n$ models produce very similar viscous corrections to the elliptic flow when $T_\zeta = 220$ MeV. Clear distinctions between these models appear only for large transverse momenta $p_T > 2$ GeV.

When $T_\zeta$ is lowered to 160 MeV (see Fig.~\ref{v2_large_bulk}), the much larger bulk viscous pressure on the space-like part of the hypersurface results in the divergence of the linearized bulk viscous corrections to $v_2(p_T)$ at intermediate values of $p_T{\,\sim\,}1.5{-}2.5$ GeV, for all three particle species considered, reflecting the corresponding sign change of their azimuthally averaged $p_T$ spectra \eqref{eq:pT_spectra}. The shear viscous corrections help offset this effect by flattening the slope of the $p_T$ spectra, but the pion and kaon elliptic flows still diverge, albeit now at slightly higher $p_T$ (as does the proton $v_2(p_T)$ from the 14--moment approximation). For realistic event-by-event simulations where the hadrons are Monte Carlo sampled from the hypersurface prior to the afterburner phase, these divergences are removed by enforcing the regulation $f_{\eq,n} + \delta f_n \geq 0$ in the Cooper--Frye formula~\cite{Shen:2014vra,McNelis:2019auj}.

In the PTM distribution, on the other hand, the bulk viscous modifications prevent the $p_T$ spectra from turning negative, and the resulting elliptic flow curves are well behaved even at high $p_T$. After including the shear stress modification, the PTM differential elliptic flows decrease but remain above the \textit{ideal} curves since the bulk viscous pressure overwhelms the shear stress on the hypersurface. The PTB elliptic flow coefficients also stay finite at high $p_T$ but are significantly larger than those computed with the PTM distribution. 

\begin{figure*}[!htbp]
\centering
\includegraphics[width=\linewidth]{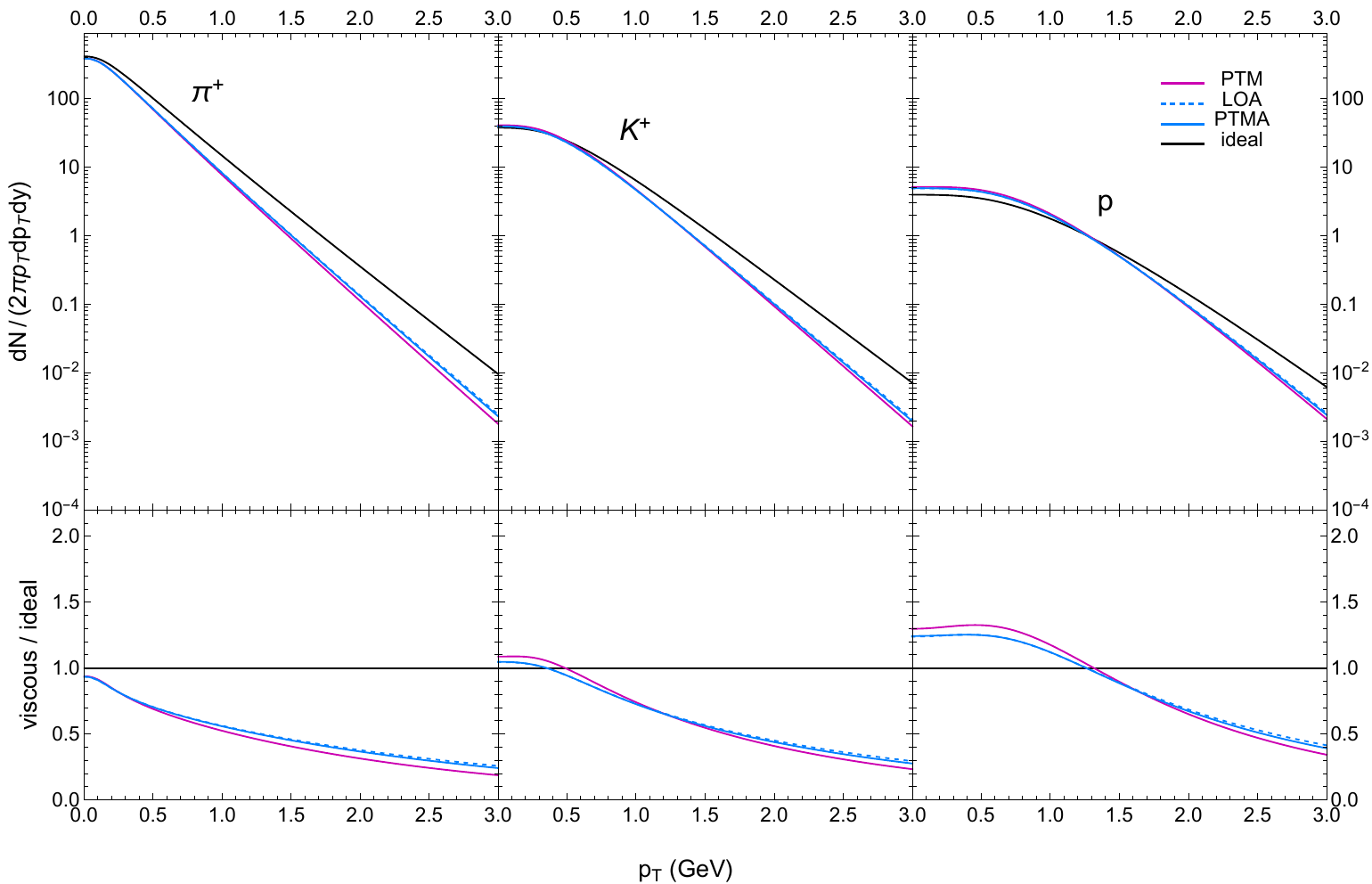}
\caption{
    The azimuthally-averaged transverse momentum spectra of ($\pi^+,K^+,p)$ (top panels) for the central Pb+Pb collision with $T_\zeta = 160$ MeV. We compare the full PTMA distribution (solid light blue) to the PTM distribution with shear and bulk modifications (solid purple), the leading order anisotropic (LOA) distribution \eqref{eq:fa} without residual shear corrections (dotted light blue), and the ideal (local thermal equilibrium) distribution (solid black). The bottom panels show the ratio of the particle spectra with viscous corrections to the ideal spectra.
\label{dN_dpT_famod}
}
\includegraphics[width=\linewidth]{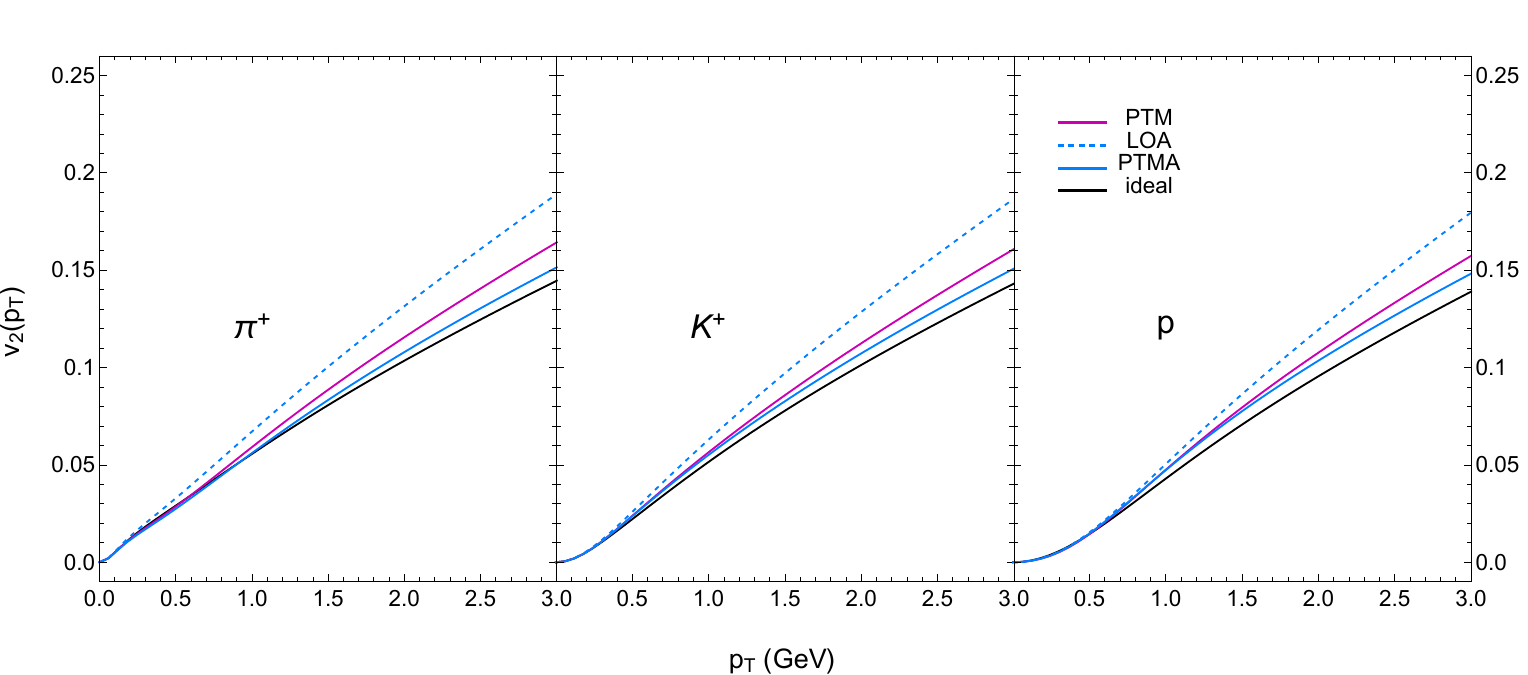}
\caption{
    The $p_T$--differential elliptic flow of ($\pi^+,K^+,p)$  for the non-central Pb+Pb collision with $T_\zeta = 160$ MeV. Similar to Fig.~\ref{dN_dpT_famod}, we compare the PTMA distribution (solid light blue) to the PTM distribution (solid purple) and leading order anisotropic (LOA) distribution (dotted light blue).
\label{v2_famod}
}
\end{figure*}
\subsection{Modified anisotropic distribution}

We close this section by looking at the particle $p_T$ spectra and differential elliptic flows computed with the PTMA distribution \eqref{eq:famod1}. Figure~\ref{dN_dpT_famod} shows the resulting transverse momentum spectra for the same central Pb+Pb collision as in Fig.~\ref{dNdpT_large_bulk}. Compared to the PTM spectra, which here include both shear and bulk viscous corrections, the PTMA kaon and proton distributions have a slightly higher mean transverse momentum, indicated by corresponding shifts in the slope and shoulder, and the pion yield slightly increases. We know from Fig.~\ref{hydro_output} that the PTM distribution underpredicts the input isotropic pressure $\Peq + \Pi$ because the bulk viscous modifications to the effective temperature and isotropic momentum space in Eq.~\eqref{eq:feqmod_2} do not perfectly reproduce the input bulk viscous pressure. The anisotropic parameters ($\Lambda, \alpha_\perp,\alpha_L$) in the PTMA distribution are optimized to correct for these errors, outputting a slightly larger isotropic pressure than the PTM distribution. One also notices that the PTMA spectra are virtually identical to the ones computed with leading order anisotropic distribution~\eqref{eq:fa}, which excludes the residual shear corrections in Eq.~\eqref{eq:Bij}. Due to the approximate azimuthal symmetry of the hypersurface, we expect the transverse shear stress $\pi_\perp^\munu$ ($\Wperp^\mu = 0$ by longitudinal boost-invariance) to have little to no impact on the azimuthally-averaged particle spectra. 

Figure~\ref{v2_famod} shows the elliptic flow coefficients from the PTMA distribution for the same non-central Pb+Pb collision as in Fig.~\ref{v2_large_bulk}. Although the PTM and PTMA models yield very similar results, the smaller bulk viscous modification in the PTMA distribution slightly brings down the $v_2(p_T)$ curves. Relative to the leading order anisotropic distribution, we see that the transverse shear corrections play a larger role in non-central collisions, damping the anisotropic flow.

\section{Conclusions}
\label{sec6}

In this work we proposed and studied a positive definite hadronic distribution for Cooper--Frye particlization that improves upon the Pratt--Torrieri distribution introduced in Ref.~\cite{Pratt:2010jt}. At an intermediate stage of the derivation, we used the RTA Boltzmann equation to identify the non-equilibrium corrections to the Boltzmann factor. This ensures that in the limit of small dissipative flows our ``modified equilibrium" distribution function reduces exactly to the first-order RTA Chapman--Enskog expansion. Even with these dissipative modifications, the distribution function remains positive definite for arbitrarily large momenta, but there is a trade-off: different from the first-order RTA Chapman--Enskog expansion, the modified equilibrium distribution no longer matches the input energy-momentum tensor and net baryon current exactly. The mismatch becomes significant already for moderately large viscous corrections. 

To minimize these errors, in Sec.~\ref{sec3b} we slightly restructured the dissipative perturbations by linearizing the viscous corrections to the momentum scales of the local-equilibrium distribution as well as their contributions to the corresponding particle yields. The resulting PTM modified equilibrium distribution bears close resemblance to the Pratt--Torrieri distribution \cite{Pratt:2010jt} but it can better reproduce the input hydrodynamic quantities in freeze-out cells subject to moderately large bulk viscous pressures. Although the output of the energy-momentum tensor calculated with the modified equilibrium distribution does not perfectly match the input $T^{\mu\nu}$, one can further improve it by applying the same technique to modify not the {\it isotropic} local-equilibrium distribution but instead the leading order {\it anisotropic} distribution (which accounts for some large dissipative effects non-perturbatively). At this point this PTMA distribution has only been constructed for systems with zero net baryon density; its generalization to nonzero baryon chemical potential will be left to future work.    

We also compared the PTM distribution to the linearized 14--moment approximation and the first-order RTA Chapman--Enskog expansion, by using the Cooper--Frye formula to compute the momentum spectra and $p_T$--differential elliptic flow coefficients of hadrons emitted from a particlization hypersurface. For small viscous corrections the azimuthally averaged transverse momentum spectra and $p_T$--differential elliptic flows generated by all of these $\delta f_n$ models are very similar. When the dissipative flows on the hypersurface (in our case mostly the bulk viscous pressure) become moderately large, the transverse momentum spectra of the linearized $\delta f_n$ corrections turn negative at intermediate $p_T$ values while those of the PTM and PTMA distributions remain positive at all $p_T$, by construction. This simultaneously prevents the $p_T$--differential elliptic flow coefficients from developing singularities caused by zero crossings of the $p_T$ spectra, a problem that plagues the linearized viscous corrections whenever the hypersurface features large bulk viscous pressures.

In practical applications the hydrodynamic evolution of the fluid inside the particlization hypersurface must be complemented by a microscopic kinetic evolution outside that surface, typically simulated by a hadronic cascade that allows for the decay of unstable resonances and their regeneration via hadronic rescattering (such as e.g. {\sc URQMD} \cite{Bleicher:1999xi} or {\sc SMASH} \cite{Weil:2016zrk}). Initialization of this cascade requires the sampling of hadron positions and momenta from the Cooper--Frye formula using Monte Carlo techniques \cite{Shen:2014lye}, rather than computing continuous particle spectra by numerically performing the Cooper--Frye integral as we have done in this work. Monte Carlo sampling requires a positive definite hadron distribution function $f_n = f_{\eq,n} + \delta f_n$. Positivity during the particle sampling process can be ensured by appropriate regularization prescriptions \cite{Shen:2014lye,McNelis:2019auj}. Combined with fluctuations from finite sampling statistics, it then becomes difficult to distinguish the modified equilibrium distributions from the regulated linearized $\delta f_n$ corrections at large momenta where few particles are emitted in any given collision event. The differences that matter in practice are those that are visible at soft momenta $p_T\lesssim 1$ GeV/$c$.

The Monte Carlo sampling of hadrons from the Cooper--Frye formula using one of the five $\delta f_n$ models discussed in this work has already been implemented in the code {\sc iS3D} \cite{McNelis:2019auj}. In addition to the PTM and PTMA distributions, it may be insightful to consider other $\delta f_n$ correction candidates, in order to fully quantify the theoretical uncertainties associated with the particlization process in heavy-ion collisions. A recent model-to-data analysis that utilized our code to reduce model selection bias in the phenomenological constraints on the quark-gluon plasma's transport properties was reported in Refs.~\cite{Everett:2020xug,Everett:2020yty}. The only $\delta f_n$ model discussed here that was not used in that analysis is the PTMA modified anisotropic distribution, which is the latest addition to the {\sc iS3D} particle sampler.

An interesting recent development \cite{Everett:2021ulz} is the proposal of a new type of positive definite particle distribution that introduces dissipative corrections to the local thermal equilibrium distribution by maximizing the entropy while using additional Lagrange multipliers to keep {\it all} ten components of the energy-momentum tensor $T^{\mu\nu}$ (including the bulk and shear viscous stresses) fixed exactly at their hydrodynamic values. The conceptual advantage of this approach is that it does not rely on any uncontrolled assumptions about the microscopic dynamics of the distribution function but uses only the available macroscopic information encoded in the hydrodynamic output for $T^{\mu\nu}$. We note that this so-called ``maximum entropy" distribution \cite{Everett:2021ulz} shares many features with our PTM and PTMA modified equilibrium distributions. It would be interesting to see whether a deeper connection exists between these two approaches, and to compare their particle spectra in event-by-event simulations of heavy-ion collisions with fluctuating initial conditions. Such studies will become feasible once a full numerical calculation of the Lagrange multipliers in the maximum entropy distribution for arbitrary energy-momentum tensors becomes available.

\acknowledgements
%
The authors would like to thank Scott Pratt, Jonah Bernhard, Derek Everett and Chandrodoy Chattopadhyay for useful discussions. This work was supported by the National Science Foundation (NSF) within the framework of the JETSCAPE Collaboration under Award No. ACI-1550223. Additional partial support by the U.S. Department of Energy (DOE), Office of Science, Office for Nuclear Physics under Award No. DE-SC0004286 and within the framework of the BEST and JET Collaborations is also acknowledged.

\appendix
\section{Moments of the distribution function}
\label{app:integrals}

We define the thermal, isotropic and anisotropic integrals that appear in this paper. For each species $n$ let
\be
  J_{kq,n} \equiv \int_p \frac{(\up)^{k-2q} ({-}p{\,\cdot\,}\Delta{\,\cdot\,}p)^q}{(2q{+}1)!!} f_{\eq,n} \bar{f}_{\eq,n}\,.
\ee
The {\it thermal integrals} over the local-equilibrium distribution are then given by
\bs
\begin{align}
  \mathcal{J}_{kq} &= \sum_n J_{kq,n}\,,
\\
  \mathcal{N}_{kq} &= \sum_n b_n J_{kq,n}\,,
\\
  \mathcal{M}_{kq} &= \sum_n b^2_n J_{kq,n}\,,
\\
  \mathcal{A}_{kq} &= \sum_n m_n^2 J_{kq,n}\,,
\\
  \mathcal{B}_{kq} &= \sum_n b_n m_n^2 J_{kq,n}\,.
\end{align}
\es
The {\it isotropic integrals} over the PTB distribution \eqref{eq:Jonah} without shear stress modifications (i.e. for $\pi_{ij} = 0$) are
\be
  \mathcal{L}_{kq} = \sum_n \int_p \frac{(\up)^{k-2q} ({-}p{\,\cdot\,}\Delta{\,\cdot\,}p)^q}{(2q{+}1)!!} f_{\lambda,n}\,,
\ee
where 
\be
  f_{\lambda,n} = \frac{g_n(1{+}\lambda_\Pi)^{-3}}{\exp\left[\dfrac{1}{T}\sqrt{m_n^2 - \dfrac{p {\,\cdot\,} \Delta {\,\cdot\,} p}{(1{+}\lambda_\Pi)^2}}\right]+\Theta_n}\,.
\ee
In particular, the modified energy density and isotropic pressure are given by $\ene^\prime(\lambda_\Pi,z_\Pi,T) = z_\Pi \mathcal{L}_{20}(\lambda_\Pi,T)$ and $\mathcal{P}^\prime(\lambda_\Pi,z_\Pi,T) = z_\Pi \mathcal{L}_{21}(\lambda_\Pi,T)$. 

\noindent Finally, the {\it anisotropic integrals} are given by 
\be
\begin{split}
  \J_{krqs} = & \sum_n  \frac{1}{(2q)!!} \int_p (\up)^{k{-}r{-}2q} (- Z {\,\cdot\,} p)^r 
\\
  &\ \  \times ({-}p {\,\cdot\,} \Xi {\,\cdot\,} p)^q \pOp^{s/2} f_{a,n} \bar{f}_{a,n}\,.
\end{split}
\ee

\section{Conservation laws in first-order approximation}
\label{app:conservation}

Here we derive the expression for the time derivatives~\eqref{eq15} by making a first-order approximation to the conversation laws~\eqref{eq:conservation_laws}. The conservation equations for the net baryon number and energy up to first order in gradients are
\be
  \dot{n}_B = - n_B \theta,\qquad
  \dot \ene = - (\ene{+}\Peq)\theta\,.
\ee
Taking the time derivative of the kinetic definitions for the net baryon and energy densities,
\be
\!\!
  n_B = \sum_n b_n \int_p (\up) f_{\eq,n}, \quad 
  \ene = \sum_n \int_p (\up)^2 f_{\eq,n},\!\!
\ee
one obtains
\be
  \dot\alpha_B = \mathcal{G}(T,\alpha_B) \theta\,, 
  \qquad
  \dot T = \mathcal{F}(T,\alpha_B) \theta \,.
\ee
The coefficients $\mathcal{G}$ and $\mathcal{F}$ appear in the RTA Chapman--Enskog expansion \eqref{eq:Chapman_Enskog}:
\bs
\begin{align}
  \mathcal G &= T\left(\frac{(\ene{+}\Peq)\N_{20} - n_B \J_{30}}{\J_{30} \mathcal{M}_{10} - \N_{20}^2}\right) \,,
\\
  \mathcal F &= T^2\left(\frac{n_B\N_{20} - (\ene{+}\Peq)\mathcal{M}_{10}}{\J_{30} \mathcal{M}_{10} - \N_{20}^2}\right) \,.
\end{align}
\es
The evolution equation for the fluid velocity up to first-order in gradients is given by
\be
\label{eq:fluid_velocity}
  \dot{u}^\mu = \frac{\nabla^\mu \Peq}{\ene{+}\Peq}\,.
\ee
After computing the spatial gradient of the equilibrium pressure $\Peq = \frac{1}{3} \sum_n \int_p (-p{\,\cdot\,} \Delta{\,\cdot\,}p) f_{\eq,n}$, Eq.~\eqref{eq:fluid_velocity} can be rewritten as
\be
\dot u^\mu = \nabla^\mu {\ln}T + \frac{n_B T}{\ene {+} \Peq} \nabla^\mu \alpha_B\,,
\ee
where we made use of the identities~\cite{Jaiswal:2014isa} $\mathcal{J}_{31} = (\ene{+}\Peq)T$ and $\mathcal{N}_{21} = n_B T$.

\bibliography{modified}

\end{document}